\documentclass[acmsmall,screen]{acmart}
\AtBeginDocument{%
  }

\setcopyright{acmlicensed}
\copyrightyear{2026}
\acmYear{2026}
\acmDOI{XXXXXXX.XXXXXXX}

\begin{document}

% \title{Inferring Time Zone from Online Community Activity}
%ITZ-fOCA!
\title{Reddit's Globalization over Twenty Years: Inferring Community Time Zone from Activity Timestamps}
% Reddit's Globalization over Twenty Years: Inferring Community Time Zone from Activity Timestamps

\author{Franco Della Negra}
\email{dellanegra.2211257@studenti.uniroma1.it}
% \orcid{}
\affiliation{%
  \institution{Sapienza University of Rome}
  \city{Rome}
  \country{Italy}
}

\author{Matteo Cinelli}
\email{matteo.cinelli@uniroma1.it}
\orcid{0000-0003-3899-4592}
\affiliation{%
  \institution{Sapienza University of Rome}
  \city{Rome}
  \country{Italy}
}

\author{Mattia Samory}
\email{mattia.samory@uniroma1.it}
\orcid{0000-0002-4916-8352}
\affiliation{%
  \institution{Sapienza University of Rome}
  \city{Rome}
  \country{Italy}
}

\renewcommand{\shortauthors}{Della Negra et al.}
\renewcommand{\shorttitle}{Community Time Zone Inference}

\begin{abstract}

Online communities are a global phenomenon, but assessing their actual geographical spread requires accurate and scalable measurement. We propose and evaluate methods that infer the time zone of online communities solely from their temporal activity patterns, requiring nothing beyond hourly activity counts. Grounding our approach in the well-established finding that posting rhythms encode circadian structure, we compare time-domain and frequency-domain methods against a parsimonious heuristic: that activity reaches its minimum around 4~a.m. local time. On Reddit, we show that the best-performing method is accurate to a sub-30-minute resolution, and that fewer than a thousand comments are sufficient to reach peak performance. Similarly, our heuristic almost matches the accuracy of more complex methods, recovering the correct time zone within a one-hour margin on average. This simple method correlates significantly with the actual distribution of Reddit's geographical spread; we validate its generalizability across communities organized around diverse cultural phenomena, from sports to finance, and apply it at scale to characterize the geographic evolution of Reddit from its founding to the present. Our method is portable across platforms and requires no user disclosure, making it a practical baseline for any study that must account for the geographic structure of online behavior.
\end{abstract}




\received{20 February 2007}
\received[revised]{12 March 2009}
\received[accepted]{5 June 2009}

\maketitle

\section{Introduction}

% \paragraph{Motivation}
% - studies use community affiliation or interaction as proxy for users' attributes, e.g., psychological states, political affiliation, and sociodemographics. useful to characterize users and platforms
% - geolocation is important because XXX
% big geographic sort?

A growing body of research has leveraged community affiliation and interaction patterns as proxies for user attributes, enabling the characterization of psychological states \cite{chancellor2020methods-868}, political orientations \cite{waller2021quantifying-ed5}, and sociodemographic profiles at scale \cite{tigunova2020reddust-2f1}. These approaches have proven valuable both for understanding individual users and for profiling platforms as sociotechnical systems \cite{balsamo2019firsthand-20e,eisenstein2018identifying-889,monti2023evidence-8d9,piccardi2024curious}. Among the many dimensions that shape online behavior, geographic location occupies a privileged position: the so-called "big geographic sort", i.e., the tendency of people to cluster with others who are in the same location, manifests online just as it does offline, making geolocation a key lens through which to examine platform dynamics \cite{bozarth2023role-092}.

% \paragraph{Gap}
% - little focus on geographic location
% - often only few communities have ground truth, limited validation of other communities
Yet despite its importance, geographic location has received comparatively little attention in the computational social science literature on online communities. Where it has been studied, ground-truth labels are typically available for only a small subset of communities, leaving the broader validity of geographic inference methods largely unvalidated.

% \paragraph{Novelty}
% - inferral from activity signals: hard to counterfeit, cheap to process, portable method across platforms as it does not depend on particular affordances. 
% Research questions:
% \begin{itemize}
%     \item[ RQ1:] How reliable is time zone inference based on time series data?
%         \item [RQ2:] How does inference quality depend on data availability?
%     \item[RQ3:] How does at-scale time-zone inference 

% \end{itemize}
This work addresses that gap by proposing and evaluating a method for inferring the geographic location of online communities (specifically, their time zone) directly from activity signals. This approach carries several methodological virtues: activity signals are difficult to counterfeit, inexpensive to compute, and, crucially, portable across platforms, as they do not depend on any platform-specific affordances or metadata. 

We organize our investigation around three research questions: (RQ1) How reliable is time zone inference based on time series data? (RQ2) How does inference quality vary with data availability? (RQ3) What is Reddit's geographical distribution, and how did it evolve?

% \paragraph{Contribution}
% - systematic comparison of the performance and computational trade-offs of multiple methods. we can infer the time zone of a subreddit within a hour margin on average, using only activity counts. We develop a simple heuristic---assuming that the time of day with the least activity is 4 a.m.---that maintains this performance. Observing the timings of a few thousand comments (or less than a month of activity) is enough to reach 95\% of this performance.    
% - extensive quantitative and qualitative evaluation of the generalizability of the method to unseen subreddits. we show that our method is able to retrieve subreddits with salient  cultural phenomena  across multiple categories, spanning sports to celebrities to politics
% - application to an at-scale characterization of time zone composition of reddit over time . we show that reddit became less US-centric and more global over time, but with specific skews: activity has a second peak around European time zones, with little adoption in Asiatic and Oceanic time zones  

Our contribution is threefold. First, we conduct a systematic comparison of multiple inference methods, evaluating both their accuracy and computational trade-offs. We demonstrate that the time zone of a subreddit can be inferred to within a one-hour margin, on average, using only activity counts, and we introduce a simple but effective heuristic that performs competitively with more computationally demanding methods: anchoring the inference on the assumption that activity is minimized around 4 a.m. local time. Importantly, observing only a few thousand comments, or less than one month of activity, suffices to recover 95\% of the maximum achievable performance. Second, we provide extensive quantitative and qualitative evaluation of the method's generalizability to unseen communities, showing that it successfully retrieves subreddits organized around salient cultural phenomena spanning sports, food, and finance. Third, we apply the method at scale to characterize the time zone composition of Reddit, showing that it correlates highly with independent measures of Reddit users' locations (Pearson's $r\approx 0.6$). We reveal how the platform has grown substantially less U.S.-centric and more global---albeit unevenly, with a secondary concentration around European time zones and markedly limited adoption in Asia and Oceania.

% \paragraph{Implications}
% - time series acts as baseline for behavioral studies on the platform: necessary to control for (cite munger at jqd:dm) 
% - time zone connects online to offline, allows focusing or accounting for location-specific factors

These findings carry implications that extend beyond descriptive platform analysis. %Activity time series constitute a behavioral baseline that must be controlled for in studies of user behavior, as
% Automated time zone inference enables addressing temporal confounds can systematically bias downstream analyses of user behavior . Conversely, 
Time zone inference provides a principled bridge between online behavior and offline context, enabling researchers to account for---or deliberately focus on---location-specific social, cultural, and institutional factors \cite{kates2021}.

% We compare:
% Histogram/kde KL distance wrt known subreddits/use
% Power spectrum KL distance wrt known subreddits/users
% Complex morlet transform phase alignment wrt known subreddits/users 
% Negative peak alignment to 4am (no need for reference) 

% there’s also the positive peak alignment method and combined_kl = sum of the three different KL values)
% Face validity check, we show the top held-out subreddits 
% Subreddits with the lowest-error inference per tz
% Subreddits that are the most discriminative of users with the lowest-error inference per tz
% ICC between subreddit and user tz for users in geolocated subreddits

% RQ2: How does inference quality depend on data availability?
% Elbow on MSRE wrt n comments/submissions 
% dellanegra.2211257@studenti.uniroma1.it check error converges to the mean shown in the overall statistics; add 1M and 10M bins
% Elbow on MSRE wrt n weeks of data
% RQ3: How does inference quality depend on location?
% Statistical tests of differences based on 
% Continent and country
% Granularity of location based on OSM type (e.g., Africa=continent vs. Tunisi=city)
% [Extra: application] RQ4: the big geographic sort in reddit interactions
% Based on one month of reddit, compute the homophily of the tz for all users

\section{Related Work}
We first discuss what is known about the relationship between geography and online behavior, and how researchers have tried to recover location from digital traces. We then turn to the specific signal we exploit: the temporal rhythm of online activity, which encodes where users are located. We discuss two complementary research lines: circadian analytics, which has characterized $p(\text{timestamps} \mid \text{location})$ and shown it to be richly structured, and the geolocation literature, which has recognized that temporal features are predictive of $p(\text{location} \mid \text{activity})$ yet has never evaluated them on their own terms. Our work addresses the methodological gap between circadian analytics and geolocation literature, and contributes to our understanding of the geography of online behavior by performing inference on the entire Reddit platform.

\subsection{Geography Shapes Online Behavior}

% The internet was supposed to dissolve geography. Early cyber-manifestoes proclaimed that online space constituted a sovereign realm entirely separate from the physical world: Barlow's \textit{Declaration of the Independence of Cyberspace} famously announced that your legal concepts of property, expression, identity, movement, and context do not apply to us''~\cite{barlow1996declaration}, while Negroponte declared that the migration of human activity into digital networks would render physical location almost irrelevant''~\cite{negroponte1995being}. The academic literature of the 1990s echoed these claims, theorizing cyberspace as a place where identity could be freely constructed and the constraints of the local finally escaped~\cite{turkle1995life, rheingold1993virtual}.
% It did not turn out that way. 

Far from escaping their local contexts~\cite{turkle1995life, rheingold1993virtual}, users bring their identity online: a fundamental premise of computational social science is that the digital traces left by users carry legible signals of who they are offline \cite{bail2018exposure-697,choudhury2013predicting-ac4}. Among the offline characteristics that shape online behavior, geographic location has proven structurally consequential. %: the so-called ``big geographic sort''~\cite{bozarth2023role-092}. 
As an exemplary study of the ``big geographic sort'', \citet{bozarth2023role-092} showed that the geographic composition of U.S. Reddit communities explains patterns of news circulation that platform-level algorithmic accounts had missed: interaction probability decays with physical distance even within a nominally global medium, and the community structure of the platform faithfully reproduces the cultural clustering of offline geography.

Yet, despite this recognized importance, geographic location is under-studied in the literature on online communities. Where it has been examined, studies typically rely on one of two approaches, each with structural limitations. The first---and by far the most common---is to use explicit geographic signals: GPS coordinates attached to individual posts, or user-provided location fields in profiles. Geotags offer high spatial precision, but, for instance, fewer than one percent of Twitter posts carry them~\cite{Pfeffer2023JustAD}, and platforms like Reddit do not support post-level geotagging at all. The second approach uses community affiliation or textual content as a proxy for location, either by identifying users who participate in explicitly localized subreddits~\cite{harrigian2018,alarfaj2025,berragan2023} or by applying natural language processing to extract geographic references from post content~\cite{stillman2024,hoffmann2020}. %While tractable, these methods are also circular by construction: they work precisely because they rely on the same geographic content they purport to infer, and they cannot be applied to communities whose geographic character is unknown or whose members do not disclose their location in text.

Both approaches operate at the level of individual users and leverage explicit geographic disclosure---either a geotag, a location field, or a geographic reference in text---from a comparatively small subset of users, leaving the broader geographic landscape of platforms like Reddit largely uncharacterized. The present work departs from this paradigm by targeting communities rather than individual users, and by relying on temporal activity patterns rather than any form of geographic content. This opens the possibility of geographic inference that is platform-agnostic and immune to the selection biases of geotagged and self-reported data. 

% \subsection{From Circadian Rhythms to Geographic Location}

\subsection{Circadian Analytics: Location In, Behavior Out}
The empirical foundation for our approach is a well-established literature demonstrating that \emph{when} people post online is a function of \emph{where} they live. Two streams of research---circadian analytics and geolocation---have approached this relationship from opposite directions.

The circadian analytics stream treats location as given and uses timestamps to characterize behavior. Its central finding is that posting activity encodes robust, location-conditioned temporal structure. \citet{morales2017} showed that Twitter activity exhibits characteristic 24-hour ``heartbeats''---cyclical patterns that reflect local work, commute, and leisure cycles and differ systematically across urban areas. \citet{zhou2023} extended this finding to Reddit, showing that these community-level rhythms correlate with population-level health outcomes derived from wearable devices. These studies effectively characterize $p(\text{timestamps} \mid \text{location})$---the conditional distribution of posting times given known geographic location---and establish that this distribution is rich, stable, and interpretable across platforms.

The same logic applies at the individual level. \citet{meyerson2023} developed a parametric model of individual bedtime for Reddit users: a flat rate during waking hours followed by a quadratic depletion centered on bedtime, fitted purely from UTC timestamps with known time zones. %The model successfully infers individual bedtimes, demonstrating the inferential power of timestamps as a behavioral signal even on an asynchronous platform. 
\citet{chen2018} independently established that habitual sleep midpoints can be recovered from the gaps and intensity drop-offs bracketing inactivity periods in Twitter data, with cross-country comparisons revealing systematic geographic variation in inferred sleep timing. \citet{scheffler2016} pushed further, showing that the onset of Twitter activity tracks population-level circadian phase across seasons and registers the disruption of daylight saving time transitions as a measurable shift in the collective temporal rhythm.

\subsection{Geolocation: Behavior In, Location Out}
The geolocation stream attempts the inverse of circadian analytics---predicting location from observable signals. Frameworks for Twitter home-location inference combine textual content (e.g., place names, dialectal vocabulary, gazetteer lookups), social network structure, and behavioral features including temporal patterns, in hierarchical ensemble classifiers that first resolve time zone and region before narrowing to the city level~\cite{mahmud2012,mahmud2014}. In these models, temporal features appear alongside content and network signals, and their standalone predictive performance is never reported. Similarly, \citet{ghoorchian2018} introduced time-slot-based partitioning to handle multi-location users---recognizing that location predictability is itself a function of time of day---but the temporal dimension structures network-based inference rather than replacing it. The necessity to complement the activity time series with other features likely stems from the noisiness, variability, and scarcity \textit{for user-level inference}. 

The geolocation literature on Reddit is sparse and rarely engages with temporal signals. \citet{harrigian2018} introduced the first geolocation inference approach for Reddit using text-based distant supervision, explicitly noting that Reddit's pseudonymity and the absence of native geotagging make supervised inference difficult. Subsequent work has followed suit, relying on subreddit participation~\cite{alarfaj2025,berragan2023}, text extraction~\cite{stillman2024}, or hybrid approaches. Across this entire body of work, location inference is treated as a content analysis problem, and activity time series are not examined as the sole input signal.

\subsection{The Gap This Paper Fills}
% Whereas the circadian literature has characterized $p(\text{timestamps} \mid \text{location})$ and shown it to be richly structured, yet has never inverted it; the geolocation literature has recognized that temporal features are predictive yet has never isolated them or evaluated them on their own terms. 
To the best of our knowledge, no existing work implements a model that takes only activity time series as input---excluding text, network, and geotag signals---and outputs a geographic location. The only exception is work by \citeauthor{la2019nationality}, which infers the time zone of dark web forum users based on their hour-of-day activity histograms \cite{la2019nationality}. However, they offer a limited qualitative evaluation of their method, applied only to four forums.

Unlike prior work, this paper inverts the circadian signal in its most general form: using only hourly activity counts, aggregated at the community level, with no text, no network, no metadata, and no platform-specific affordances beyond the timestamp itself. The method inherits the methodological virtues of the circadian literature---temporal signals are difficult to counterfeit, computationally cheap to extract, and present by definition wherever activity occurs---while addressing the inferential question that the geolocation literature has left open. By targeting communities rather than individual users, we move from a needle-in-a-haystack problem (finding the rare geotagged user) to a signal-in-aggregate problem, obtaining the statistical power that makes timestamp-only inference not merely feasible but accurate.

\section{Data}\label{sec:data}

In this section, we describe the pipeline utilized to transform raw digital traces into a reliable analytical dataset. We first describe how we acquired data and curated our geographic ground truth, before outlining the temporal preprocessing steps required to extract genuine human cyclical behaviour.

\subsection{Data Collection}\label{sec:data:collection}
We utilized data collected initially by PushShift and more recently under the Arctic Shift Project, which encompasses the entirety of Reddit from 2005 to 2024 \cite{academictorrents}. We refer to this dataset as \textbf{all\_subreddits}, which we use to extract time series data for subreddits. Table \ref{tab:dataset_statistics} summarizes the statistics of processed data.  

We complement this dataset by curating a smaller, high-quality subset of subreddits that discuss specific locations around the globe. RedditLists\footnote{\url{https://www.reddit.com/r/LocationReddits/wiki/index/}} offers a user-maintained list of such subreddits, organized around continents and states. The scope of this list is global: it includes 3309 subreddits in North America, 124 in South America, 486 in Asia, 1449 in Europe, 295 in Oceania, and 145 in Mexico, Central America, \& Caribbean. Most importantly, activity in these subreddits has been validated to reflect the actual population from the corresponding localities \cite{balsamo2019firsthand-20e,waller2021quantifying-ed5}. We refer to this subset as \textbf{location\_subreddits}. We map these subreddits to time zones (see Sec. \ref{sec:data:labeling}) and use them to measure the performance of our inference methods. 
%Furthermore, for methods that require anchor points to propagate known time zones to unlabeled subreddits, splitting labeled data appropriately to avoid information leakage and ensuring correct performance estimation (see Sec. \ref{sec:methods:setup_and_metrics}).

We further augment this dataset by retrieving categories from the overview of existing subreddits.\footnote{ At the time of data collection, the page \url{https://www.reddit.com/best/communities/} listed subreddits by popularity together with their metadata, including category. At the time of writing, the page \url{https://www.reddit.com/explore/} provides a similar per-category disaggregation.} The most popular subreddits fall under 52 consolidated categories (see Tab. \ref{tab:dataset_statistics}), spanning topics from news to internet culture. However, more niche communities are either absent from the list or report unique categories, such as Men's Health. We use the subset of 221,460 unique subreddits under the 52 common categories (accounting for 380,501 assigned topics), which we call \textbf{categorized\_subreddits}, to unpack the geographic variation of the subreddits in each topic.

\begin{table*}[t]
\centering
\caption{Summary of processed data. The top panel details the three primary dataset groupings. The bottom panel provides a full breakdown of the 221,460 subreddits distributed across the 52 consolidated categories.}
\label{tab:dataset_statistics}

\textbf{Dataset Statistics} \\
\vspace{0.1cm}
\resizebox{\columnwidth}{!}{%
\begin{tabular}{@{}lrl@{}}
\toprule
\textbf{Subset Name} & \textbf{Count ($N$)} & \textbf{Description / Filters} \\
\midrule
\texttt{all\_subreddits} & 925,156 & Full Arctic Shift Reddit archive (2005--2024). \\
\texttt{categorized\_subreddits} & 221,460 & Subset mapped to the top 52 consolidated topics. \\
\texttt{location\_subreddits (Raw)} & 5,943 & Geographically validated baseline from RedditLists. \\
\texttt{location\_subreddits (Filtered)} & 1,134 & Filtered for singular time zones, OSM classes, and feature availability. \\
\bottomrule
\end{tabular}
  }%

\vspace{0.5cm}

\textbf{Categorized Subreddits Breakdown ($N$ = 221,460)} \\
\vspace{0.1cm}
\begin{tabular}{@{}lr | lr@{}}
\toprule
\textbf{Category} & \textbf{Count} & \textbf{Category} & \textbf{Count} \\
\midrule
Gaming & 55,131 & Home and Garden & 4,243 \\
Music & 27,602 & Fashion & 3,921 \\
Technology & 25,304 & Marketplace and Deals & 3,785 \\
Place & 21,328 & Fitness and Nutrition & 3,220 \\
Internet Culture and Memes & 15,144 & Outdoors and Nature & 3,180 \\
Sports & 13,182 & Family and Relationships & 3,171 \\
Learning and Education & 12,837 & Crafts and DIY & 2,894 \\
Television & 12,519 & Meta/Reddit & 2,592 \\
Animals and Pets & 10,977 & History & 2,464 \\
Podcasts and Streamers & 10,269 & Beauty and Makeup & 2,293 \\
Hobbies & 10,190 & Activism & 2,170 \\
Art & 10,139 & Ethics and Philosophy & 1,867 \\
Cars and Motor Vehicles & 9,553 & Travel & 1,801 \\
Anime & 9,121 & Culture, Race, and Ethnicity & 1,702 \\
Reading, Writing, and Literature & 8,799 & World News & 1,191 \\
Business, Economics, and Finance & 7,442 & Law & 1,173 \\
Careers & 7,167 & Gender & 1,075 \\
Mature Themes and Adult Content & 7,078 & Military & 819 \\
Celebrity & 6,817 & Sexual Orientation & 766 \\
Food and Drink & 6,649 & Women's Health & 534 \\
Medical and Mental Health & 6,551 & Trauma Support & 525 \\
Tabletop Games & 6,014 & Addiction Support & 331 \\
Crypto & 5,631 & Men's Health & 232 \\
Programming & 5,585 & & \\
Funny/Humor & 5,263 & & \\
Movies & 4,752 & & \\
Religion and Spirituality & 4,555 & & \\
Politics & 4,544 & & \\
Science & 4,409 & \textit{Other (Excluded)} & \textit{703,696} \\
\bottomrule
\end{tabular}
\end{table*}

\subsection{Geolocation: Ground-Truth Subreddit Time Zone} \label{sec:data:labeling}

Next, we establish a mapping between subreddits and their physical time zones. %This process necessitates formalizing explicit geographic assumptions and applying a strict curation pipeline to prevent data contamination. The fundamental assumption at the community level is that there is a direct correspondence between a localized subreddit (e.g., a community dedicated to a specific city or region) and the physical geographic boundaries of that location. 
To operationalize this, geolocate subreddits in the \texttt{location\_subreddits} dataset. We leverage RedditLists' taxonomy, which includes the Nation/State and name of a location, to compose descriptive strings that we pass to the Nominatim geocoder service. Then, we use the \texttt{timezonefinder} Python package to attribute time zones based on the inferred latitude and longitude of each location. To ensure these reference signals are pristine, we further apply three filters:

\begin{itemize}
    \item \textit{Geographic Specificity:} The ground truth set is restricted to subreddits representing definitive geographic entities by filtering specifically for OpenStreetMap classes denoting administrative or physical regions, namely: \textit{boundary}, \textit{place}, and \textit{natural}. Other classes are explicitly excluded as they are likely to introduce location errors.
    
    \item \textit{Single Time Zone:} We explicitly remove any subreddits whose geographic bounding boxes span multiple time zones. This ensures that every baseline subreddit is associated with a maximum of one time zone.
    
    \item \textit{Minimum Class Representation:} Before any inference begins, the algorithm counts the frequency of each time zone in the dataset. Any time zone that contains only a single subreddit is discarded, as it is required to ensure at least one reference instance per class during the cross-validation framework: as will be seen later, a time zone must have at least two representatives so that one can serve as the target while the other remains in the reference pool.
\end{itemize}

\subsection{Time-Series Extraction and Preprocessing}

Raw social media data is notoriously noisy. Before extracting temporal features, the raw hourly activity time series (e.g., counts of comments) for the selected subreddits must undergo a data preprocessing pipeline to isolate the underlying daily rhythms from high-frequency noise, long-term trends, and extreme viral events. 

\subsubsection{Sparsity Filtering} 
Reliable rhythmic patterns cannot be inferred from extremely sparse time series. Time series with fewer than 50 non-zero hourly observations are excluded to guarantee statistical reliability. 
% Furthermore, to prevent artificial skewing from periods of inactivity at the end of the data collection, time series are capped to their last non-zero observation. The pipeline optionally restricts the analysis window to the last $N$ days or comments, retaining only the most relevant temporal data.

\subsubsection{Log Transformation} 
Online activity is prone to sudden, massive spikes, such as a breaking news event, that can dwarf standard daily interactions. To stabilize variance and reduce the impact of these extreme spikes, a logarithmic transformation is applied to the raw counts. 
% Extreme anomalies are then identified using a standard Z-score threshold (e.g., 3 standard deviations). Rather than excluding these outliers and creating artificial gaps in the timeline, their values are capped at the highest non-outlier value present in the dataset.

\subsubsection{Temporal Regularization} 
Then, we regularize the data by binning the continuous timestamps into discrete 1-hour intervals, treating missing data as zeros --- as they represent a genuine absence of activity. This yields the continuous, uniformly spaced raw time series $X = \{x_t\}_{t=1}^T$.

\subsubsection{Detrending} 
Beyond short-term sparsity, subreddit activity is highly susceptible to macro-level shifts, such as gradual community growth, viral spikes, or steady decline. To isolate the underlying 24-hour cyclical behavior from these long-term trends, we apply a smoothing procedure to estimate the local trend $\mu_t$. Rather than utilizing a standard rectangular moving average---which abruptly truncates data and can introduce artificial "ringing" artifacts in the frequency domain---we compute a centered moving average using a Hann window spanning $W = 384$ hours (16 days). The Hann window smoothly tapers the weights $w_k$ toward the edges of the observation window:
\begin{displaymath}
   \mu_t = \frac{\sum_{k=-W/2}^{W/2} w_k x_{t+k}}{\sum_{k=-W/2}^{W/2} w_k}
\end{displaymath}
\begin{displaymath}
     w_k = \frac{1}{2} \left( 1 + \cos\left(\frac{2\pi k}{W}\right) \right) \,.
\end{displaymath}
Subtracting this local trend from the original signal flattens the macro-level curve, effectively exposing the community's daily rhythmic baseline. 
%Finally, downstream continuous probability modeling (such as fitting a Poisson distribution via our Generalized Additive Model) mathematically requires strictly non-negative inputs. To satisfy this constraint without altering the amplitude or shape of the cycle, the entire detrended sequence is shifted upwards by subtracting its global minimum:
% \begin{displaymath}
%     \hat{x}_t = (x_t - \mu_t) - \min_{i \in \{1,\dots,T\}} (x_i - \mu_i)
% \end{displaymath}
% This sequence of operations produces the final regularized, detrended signal $\hat{X} = \{\hat{x}_t\}_{t=1}^T$, cleanly isolating the temporal rhythm for feature extraction.

\section{Time Zone Inference Methods}

% The internet operates 24/7, but humans do not. Despite the seemingly borderless nature of online communities, human biology and local societal norms dictate when we sleep, work, and post online. Consider a local city forum on Reddit: even if it is accessible globally, the bulk of its interactions will inevitably ebb and flow with the rising and setting of the sun in its specific geographic location. 

\begin{figure}
    \centering
    \includegraphics[width=\textwidth, trim=0 20 0 20, clip]{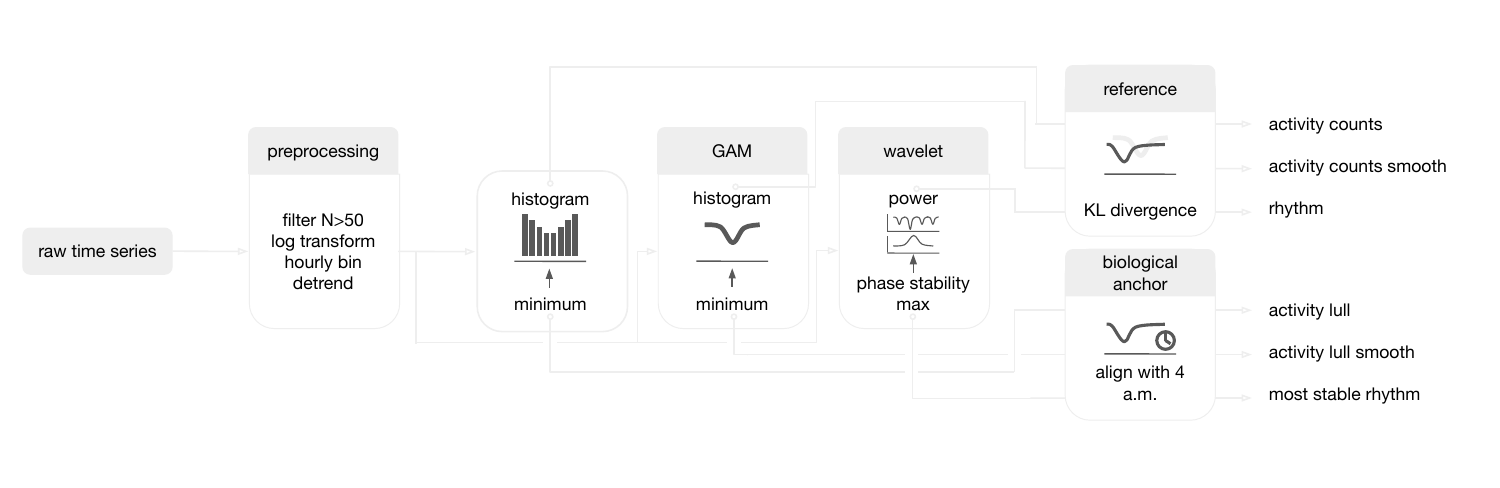}
    \caption{Overview of the processing pipeline for the six inference methods compared in this work.}
    \label{fig:pipeline}
\end{figure}

In this section, we present a methodological framework to answer a central question: Can we accurately infer a community's geographic time zone solely from its temporal activity footprint? Our model bridges the gap between raw, noisy digital footprints and localized temporal patterns. We achieve this through a three-step architecture: the circadian feature extraction, the definition of inference strategies, and a rigorous cyclical evaluation phase. Figure \ref{fig:pipeline} provides an overview for the computational pipeline for time zone inference.

\subsection{Circadian Feature Extraction}
Following preprocessing, distinct representations of the 24-hour activity cycle are extracted to capture different facets of the underlying circadian patterns, spanning both the time and frequency domains.

\subsubsection{Time-Domain Features}

\paragraph{Normalized Activity Histograms:} The most direct representation of activity patterns is the shape of the activity histogram \cite{la2019nationality,piccardi2024curious}. Preprocessed counts are aggregated by the hour of the day (0-23 UTC) and normalized to form a discrete probability distribution:
\begin{displaymath}
    P(h) = \frac{c_h}{\sum_{j=0}^{23} c_j}, \quad \text{for } h \in \{0, 1, \dots, 23\}
\end{displaymath}
From this, we extract the hour corresponding to the absolute minimum activity:
\begin{displaymath}
    h_{\text{min}} = \arg\min_{h} P(h)
\end{displaymath}

\paragraph{Cyclic Smoothing:} Since raw histograms can be sparse, we apply smoothing techniques. We use a Generalized Additive Model (GAM), fitting a Poisson distribution to the hourly aggregate data via a cyclic cubic regression spline:
\begin{displaymath}
    \log(\lambda_h) = \beta_0 + \sum_{j=1}^{k} \beta_j b_j(h)
\end{displaymath}
where $\lambda_h$ is the expected count at hour $h$, and $b_j(h)$ are the cyclic basis functions. This cyclic constraint and the cubic spline ensure a smooth, continuous transition from hour 23 back to hour 0. We extract the hour of minimum activity from this smoothed curve:
\begin{displaymath}
    h_{\text{smooth\_min}} = \arg\min_{h \in [0, 24)} \lambda_h
\end{displaymath}

\subsubsection{Time-Frequency Domain Features}

To measure the intensity and phase of the 24-hour rhythm---rather than just activity volume---we apply a Continuous Wavelet Transform, capturing both its spectral properties and temporal localization \cite{oliveiranoyearspatio-temporal-663}.

\paragraph{Wavelet Choice:} The transformation utilizes a complex Morlet wavelet (\texttt{cmor1.5-1.0}), which allows us to capture both frequency- and time-domain information simultaneously:
\begin{displaymath}
    \psi(t) = \frac{1}{\sqrt{\pi f_b}} e^{i 2 \pi f_c t} e^{-t^2 / f_b}
\end{displaymath}
where $f_c = 1.0$ is the center frequency and $f_b = 1.5$ is the bandwidth parameter.

\paragraph{Scale Selection:} The wavelet scales are specifically centered around a target frequency of 24 hours. The wavelet coefficient $W_n(s)$ at time $n$ and scale $s$ is computed as:
\begin{displaymath}
    W_n(s) = \sum_{t=0}^{N-1} \hat{x}_t \psi^* \left( \frac{t-n}{s} \right)
\end{displaymath}

\paragraph{Feature Derivation:} For each point in time, the wavelet coefficients yield both the power---the squared absolute amplitude, indicating the strength of the 24-hour rhythm---and the phase---the angle, indicating the position within the cycle:
\begin{displaymath}
    \text{Power}_n = |W_n(s)|^2, \quad \text{Phase}_n = \arg(W_n(s))
\end{displaymath}

These localized wavelet metrics are then aggregated by hour of the day to calculate the mean 24-hour power, mean phase, and phase coherence, i.e., the consistency of the phase at that specific hour across different days:
\begin{displaymath}
    R_h = \left| \frac{1}{K_h} \sum_{n \in \mathcal{T}_h} e^{i  \text{Phase}_n} \right|
\end{displaymath}
where $\mathcal{T}_h$ represents the temporal indices corresponding to hour $h$, and $K_h$ is the number of observations for that hour. 
% We record the hour exhibiting the maximum phase coherence:
% \begin{displaymath}
%     h_{\text{morlet\_phase\_max}} = \arg\max_{h} R_h
% \end{displaymath}

\paragraph{Phase Stability:} %While global phase coherence captures the average alignment of the cycle, real-world rhythms can drift. 
To capture the day-to-day consistency of a community's schedule, we introduce a measure of phase stability based on consecutive variations. We calculate the phase differences ($\Delta \text{Phase}$) between consecutive daily cycles for each hour. A rhythm is considered highly stable if this day-to-day phase shift approaches zero, indicating that the community reliably executes behaviors at the same time each day. We extract the hour of maximum phase stability by isolating the time windows where the phase varies the least:
\begin{displaymath}
    h_{\text{stable\_phase}} = \arg\min_{h} \left( \mathbb{E} \left[ | \text{Phase}_{d, h} - \text{Phase}_{d-1, h} | \right] \right)
\end{displaymath}

\subsection{Time Zone Inference Strategies}
We use these extracted features to infer the geographic time zone (UTC offset, i.e., the difference in hours and minutes between Coordinated Universal Time and the standard time at a particular place) associated with a subreddit. The methodology compares two primary inference paradigms: the biological anchor and reference-based inference.

\subsubsection{Biological Anchor}
In fields heavily dependent on human alertness, such as aviation, performance is fundamentally tied to the "circadian low"---a window of peak sleepiness generally occurring between 3:00 AM and 5:00 AM. 

It is reasonable to expect that the broader, usually active population is also overwhelmingly asleep during this time frame. Building on this premise, which aligns with Figure \ref{fig:time_domain_features}(a), this method assumes a universally anchored behavioral baseline that requires no external geographic reference. For time-domain metrics, we operationalize this by assuming an activity minimum (a ``lull time") centered at 4:00 AM local time. The UTC offset is calculated by taking the circular difference between the assumed local 4:00 AM lull ($h_{\text{lull}} = 4$) and the empirically observed activity minimum in UTC ($h_{\text{obs}}$):
\begin{displaymath}
        \text{Offset}_{\text{UTC}} = \left( (h_{\text{lull}} - h_{\text{obs}} + 12) \% 24 \right) - 12
\end{displaymath}
This approach is applied individually to:
\begin{itemize}
    \item The raw histogram minimum (\texttt{Activity Lull}).
    \item The GAM-smoothed minimum (\texttt{Activity Lull Smooth}). 
    \item The hour of maximum wavelet phase stability (\texttt{Most Stable Rhythm}), as it is reasonable to assume that the most stable rhythmic phase across days inherently corresponds to the strictly enforced biological sleep period.
\end{itemize}

\subsubsection{Reference-Based Inference}

Rather than assuming a fixed behavioral anchor, this paradigm infers the target's time zone by comparing its rhythmic signature against a ground-truth reference pool of localized subreddits with known UTC offsets. This acts as a nearest-neighbour classification framework where the target inherits the geographic properties of its closest match.

The reference pool is formalized as a set of tuples coupling each reference subreddit's extracted features with its exact ground-truth UTC offset. When classifying a target subreddit, the algorithm computes a distance metric between the target's features and those of every reference in the pool. %Once the mathematical minimum distance is found, the algorithm assigns the ground-truth offset of that optimal reference to the target.
To quantify the similarity between activity patterns, we compute the Kullback-Leibler (KL) Divergence between the target probability distribution $P$ and the reference distribution $Q_r$. 
% The mathematical implementation explicitly adapts to the structural dimensionality of the extracted features:
% 
% \textit{Discrete Representations:} For the raw activity histograms and aggregated wavelet power, the features are explicitly pre-normalized during extraction into discrete probability distributions over a bounded state space of 24 hourly bins. The divergence is computed exactly as:
% \begin{displaymath}
%     D_{KL}(P \parallel Q_r) = \sum_{h=0}^{23} P(h) \log\left(\frac{P(h)}{Q_r(h)}\right)
% \end{displaymath}
% 
% \textit{Continuous Representations (GAM):} The GAM-smoothed curves provide a continuous representation of the daily rhythm. To compute the divergence, we approximate the continuous information loss integral using a high-resolution discrete Riemann sum. During extraction, the GAM predicts raw activity counts across $N = 1000$ evenly spaced intervals spanning the 24-hour cycle. These raw vectors are dynamically normalized into valid probability mass functions prior to computation. The continuous divergence is thus approximated as:
% \begin{displaymath}
%     D_{KL}(p \parallel q_r) = \int_0^{24} p(x) \log\left(\frac{p(x)}{q_r(x)}\right) dx \approx \sum_{i=1}^{1000} P(x_i) \log\left(\frac{P(x_i)}{Q_r(x_i)}\right)
% \end{displaymath}
% 
% In both cases, 
The target inherits the UTC offset of the reference subreddit that minimizes this divergence (e.g., \texttt{Rhythm}, \texttt{Activity Counts Smooth}):
\begin{displaymath}
    r^* = \arg\min_{r \in \mathcal{R}} D_{KL}(P \parallel Q_r)
\end{displaymath}

This approach is applied to:
\begin{itemize}
    \item The normalized activity histogram (\texttt{Activity Counts}).
    \item Its GAM-smoothed version (\texttt{Activity Counts Smooth}).
    \item The wavelet power spectrum (\texttt{Rhythm}).
\end{itemize}

\subsubsection{Summary of Time Zone Inference Methods}
In all, we evaluate six inference methods:
\begin{itemize}
    \item \texttt{Activity Counts}, which attributes the time zone of the reference subreddit with the smallest Kullback-Leibler divergence on the normalized activity histogram.
    \item \texttt{Activity Lull} which aligns the minimum of the normalized activity histogram at 4~a.m. in local time.
    \item \texttt{Activity Counts Smooth}, which is identical to \texttt{Activity Counts}, but enhances resolution of the time series through GAM smoothing.
    \item \texttt{Activity Lull Smooth}, which is identical to \texttt{Activity Lull}, but operates on the GAM-smoothed time series.
    \item \texttt{Rhythm}, which recovers the time zone of the reference subreddit with the closest wavelet power spectrum using Kullback-Leibler divergence.
    \item \texttt{Most Stable Rhythm}, which aligns the maximum wavelet phase stability at 4~a.m. in local time.
\end{itemize}
All methods rely on the same underlying data and preprocessing steps. From the point of view of resource demands, whereas reference-based methods (\texttt{Activity Counts}, \texttt{Activity Counts Smooth}, and \texttt{Rhythm}) require a minimal set of subreddits labeled with ground-truth time zone information and a computational step of distance-based retrieval, biological anchor methods (\texttt{Activity Lull}, \texttt{Activity Lull Smooth}, and \texttt{Most Stable Rhythm}) are applicable without any additional information and through a simple linear scan of the data. A further important practical consideration is that methods relying on wavelet transformation (\texttt{Rhythm} and \texttt{Most Stable Rhythm}) and GAM smoothing (\texttt{Activity Counts Smooth} and \texttt{Activity Lull Smooth}) incur in a one-time additional computational cost compared to the more parsimonious histogram-based methods (\texttt{Activity Counts} and \texttt{Activity Lull}).

\subsection{Evaluation Setup and Metrics}
\label{sec:methods:setup_and_metrics}

To assess the robustness and generalizability of these methods, we apply them to the set of subreddits with known locations (\texttt{location\_subreddits}). The evaluation relies on a repeated random sampling validation strategy. Finally, all model inferences are benchmarked against a probabilistic baseline.

\subsubsection{Cross-Validation Framework}
% The evaluation procedure is executed over 10 independent iterations, in which the data is partitioned into two distinct sets:
% \begin{itemize}
%     \item {The Reference Pool:} A subset of the subreddits, used for alignment in reference-based inference.
%     \item {The Target Pool:} The remaining entities whose time zones are strictly withheld and treated as unknown targets to be inferred.
% \end{itemize}

The evaluation procedure is executed over 10 independent iterations to guarantee robust results. In every iteration, 20\% of the subreddits are randomly sampled to serve as references for reference-based inference methods, while the remaining 80\% are used for evaluation. We use a stratified shuffle split strategy to ensure that the reference pool is never accidentally depleted of a minority time zone.

% Crucially, the split is strictly stratifying according to the ground-truth UTC offsets. Because of this stratification, both pools perfectly mirror the global, macro-level geographic distribution of subreddits present in the full dataset. This guarantees that the reference pool is never accidentally depleted of a minority time zone, which would preclude the reference-based algorithm from predicting it.

\subsubsection{Evaluation Metrics}
% To comprehensively quantify the performance of the various temporal alignment methods (biological anchor vs. reference-based), 
The inferred offsets are compared against the ground truth using the following suite of metrics:
% . 

\begin{itemize}
    \item \textbf{Mean Circular Error (MCE):} The primary metric is the absolute shortest temporal distance on a 24-hour clock between the predicted offset and the actual offset. 
    
    This is calculated using modulo 24 arithmetic to correctly handle the midnight boundary --- note that because time of day is cyclical, standard linear error metrics would be invalid. For an actual offset $y$ and an inferred offset $\hat{y}$, the error is defined as:
    \begin{displaymath}
        E = \min\left( |y - \hat{y}|, 24 - |y - \hat{y}| \right)
    \end{displaymath}
    
    \item \textbf{Accuracy:} A strict classification metric representing the exact match rate (where $\hat{y} = y$), meaning the algorithm predicted the exact hour perfectly.
    
    \item \textbf{Circular Correlation ($\rho$):} We compute the Jammalamadaka circular correlation coefficient to measure the strength of the directional relationship between the predicted and actual offsets \cite{jammalamadaka1988correlation}. The offsets in hours are first converted to radians ($\alpha = y \cdot \frac{2\pi}{24}$), and the circular mean is computed using trigonometric functions to establish the correlation over the periodic space.
    
    \item \textbf{Weighted Kappa ($\kappa$):} To measure inter-rater reliability while penalizing predictions based on how far away they are from the truth, we calculate a linearly weighted Cohen's Kappa. The 24 discrete hours are treated as sequential ordinal categories, ensuring that severe geographic misclassifications are penalized more heavily than minor adjacent errors.
    
    \item \textbf{Weighted F1-Score:} The inferred and actual offsets are rounded to the nearest integer to serve as discrete classes. A frequency-weighted F1-score is then applied to evaluate the balance of precision and recall,  handling the inherent class imbalance of global time zones.
\end{itemize}

\subsubsection{Baseline Comparison}
To ensure the statistical significance of our findings, all model inferences are compared against a naive stratified baseline (\textit{Dummy Classifier}). This probabilistic baseline predicts time zones randomly, but its predictions are weighted by the prior probability of the ground-truth geographic distribution found within the reference set for that specific iteration.

\section{Results}

Next, we explore the distribution of time- and frequency-domain features, before addressing the three research questions in turn: we evaluate the methods, ablate their performance based on data and time availability, and apply them to describe Reddit's geographical composition.

In more detail, Section \ref{subsec:EDA} provides an exploratory analysis of Activity Counts, specifically, activity histograms for subreddits included in the \textit{location\_subreddits} ground-truth dataset. Next, Section \ref{subsec:RQ1} addresses RQ1 by evaluating six distinct methods: \texttt{Activity Counts}, \texttt{Activity Counts Smooth}, \texttt{Activity Lull}, \texttt{Activity Lull Smooth}, \texttt{Rhythm}, and \texttt{Most Stable Rhythm}. These methods employ one of the two available inference strategies: Biological Anchor in the case of lull-based methods and Reference-Based Inference in the remaining four cases. Section \ref{subsec:RQ2} then addresses RQ2 through a robustness check, assessing methods' performance across decreasing levels of data availability. Finally, Section \ref{subsec:RQ3} builds upon these findings to evaluate the geographic distribution of Reddit communities, as specified in RQ3.

\subsection{Exploratory Data Analysis}
\label{subsec:EDA}
Before turning to inference, we examine the raw circadian features extracted from activity time series in the \textit{location\_subreddits} dataset, to establish whether the geographic signal is visible to the naked eye, and to contextualize the methodological choices outlined in the previous section.

\begin{figure}[h!]
  \centering
  
  \begin{minipage}{0.48\textwidth}
    \centering
    \includegraphics[width=\textwidth]{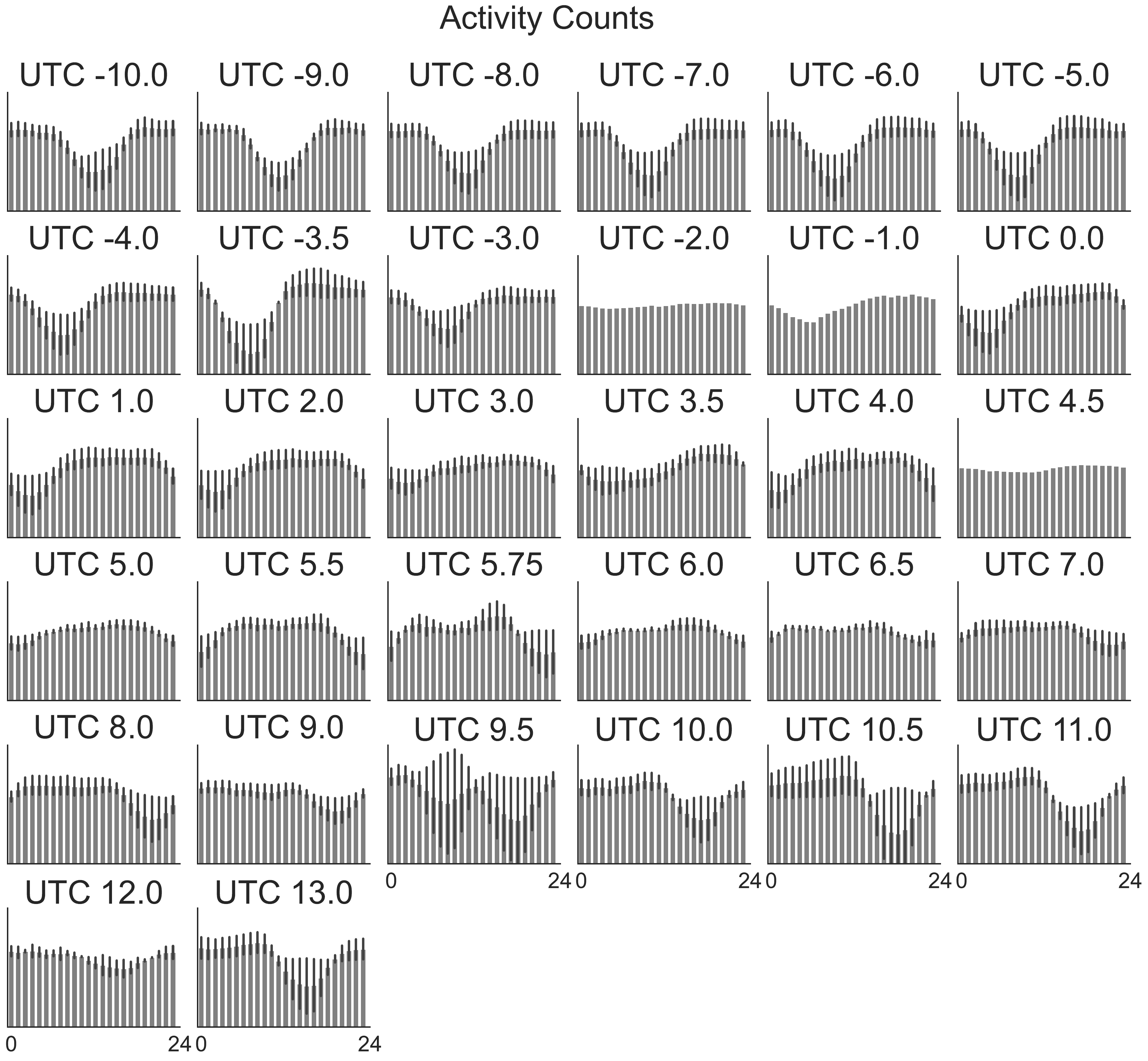}

  \end{minipage}
  \hfill
  \begin{minipage}{0.48\textwidth}
    \centering
    \includegraphics[width=\textwidth]{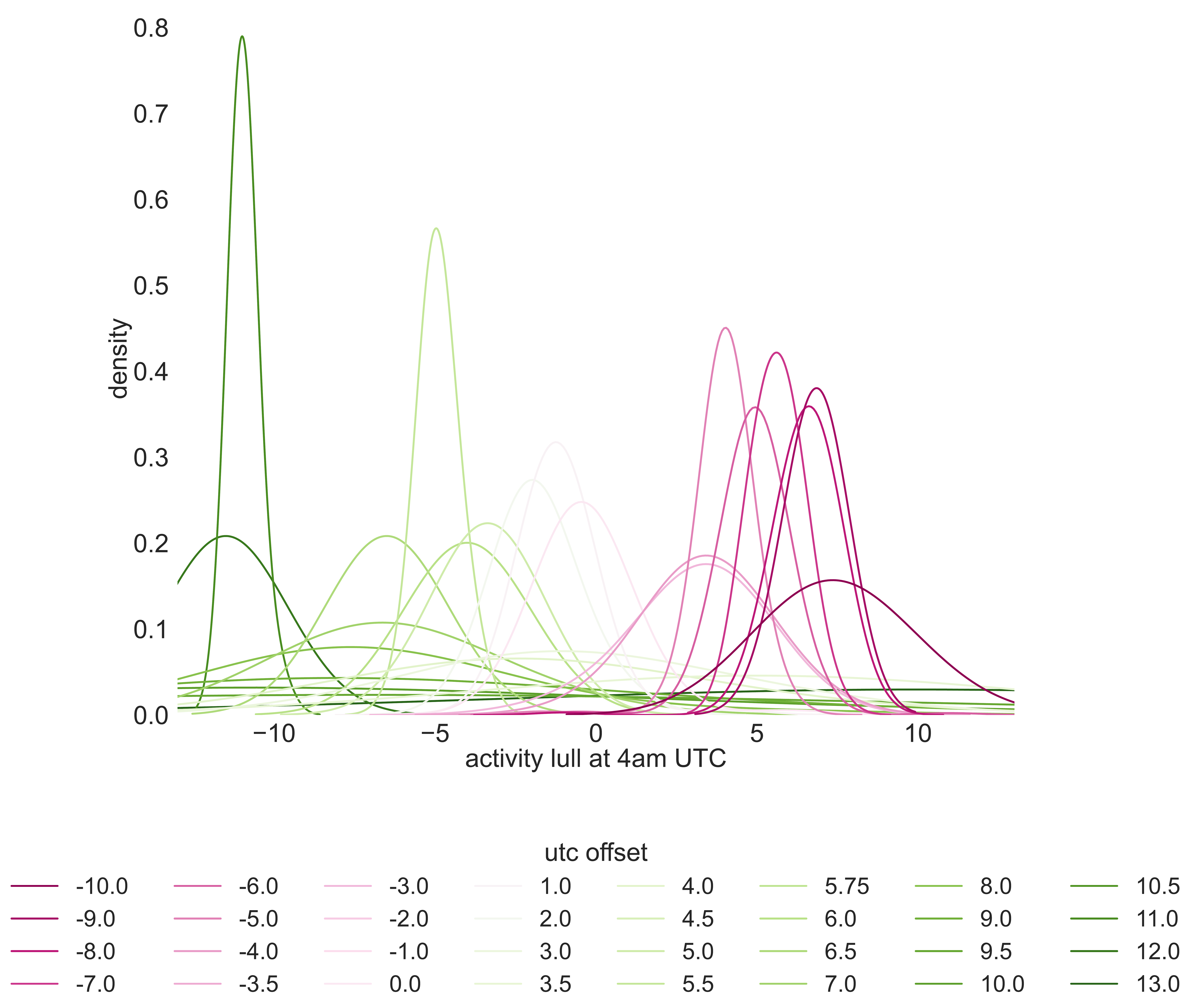}
    
  \end{minipage}
  
  \vspace{0.6cm}
  
  \caption{Time-domain feature distributions. (Left) Hourly activity distributions across different UTC offsets. The minimum activity period in the raw histograms shifts based on the relative UTC offset. (Right) Plotting the density distribution of extracted activity lulls, this hour of minimum activity appears to correspond consistently with 4:00 AM local time.}
  \Description{Two side-by-side charts. The left chart is a facet grid showing 24 bar charts of hourly activity, shifting smoothly as the time zone changes. The right chart shows overlapping bell-curve density plots highlighting the minimum activity hours.}
  \label{fig:time_domain_features}
\end{figure}

\begin{figure}[h!]
  \centering
  
  \begin{minipage}{0.48\textwidth}
    \centering
    \includegraphics[width=\textwidth]{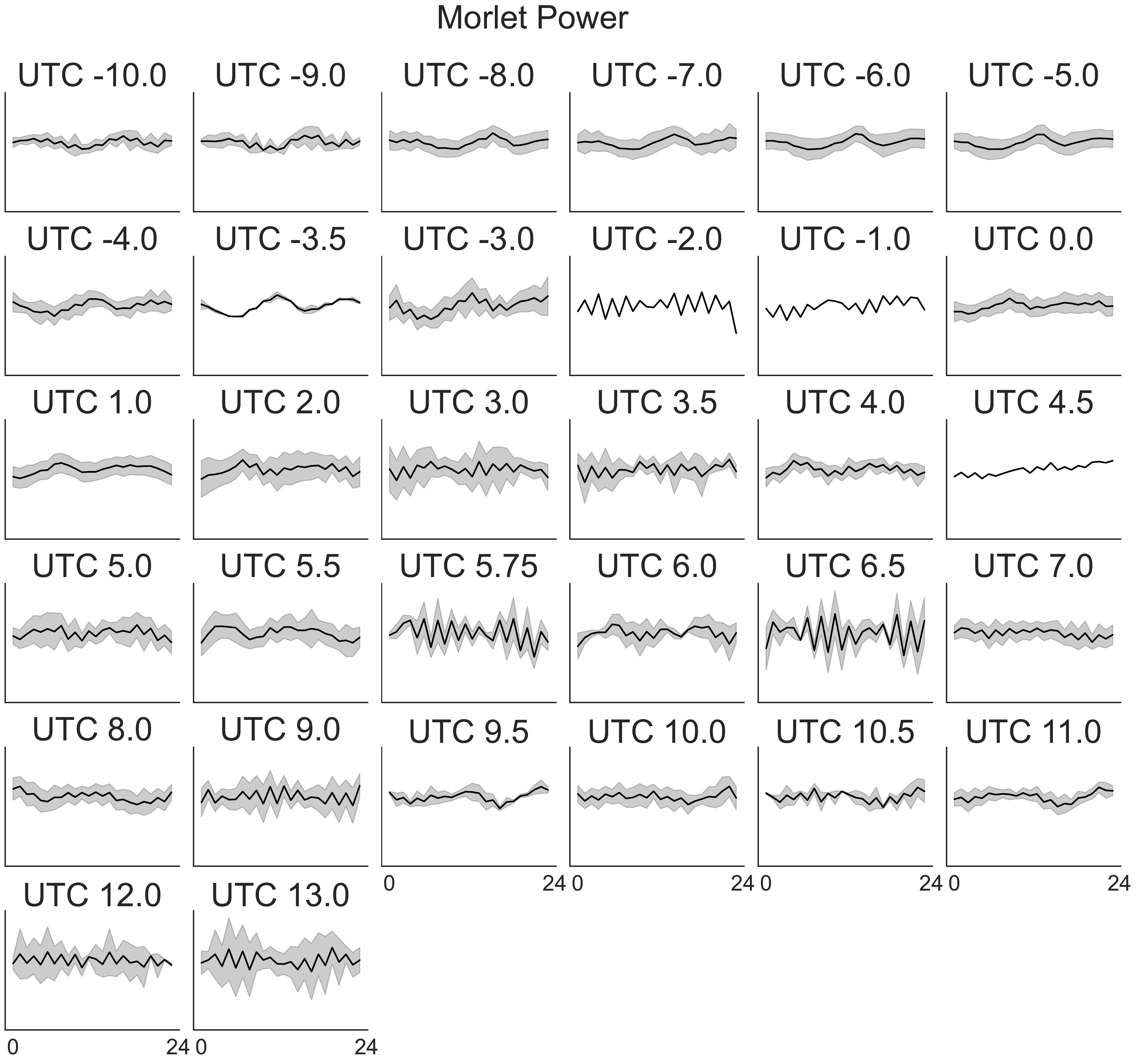}

  \end{minipage}
  \hfill
  \begin{minipage}{0.48\textwidth}
    \centering
    \includegraphics[width=\textwidth]{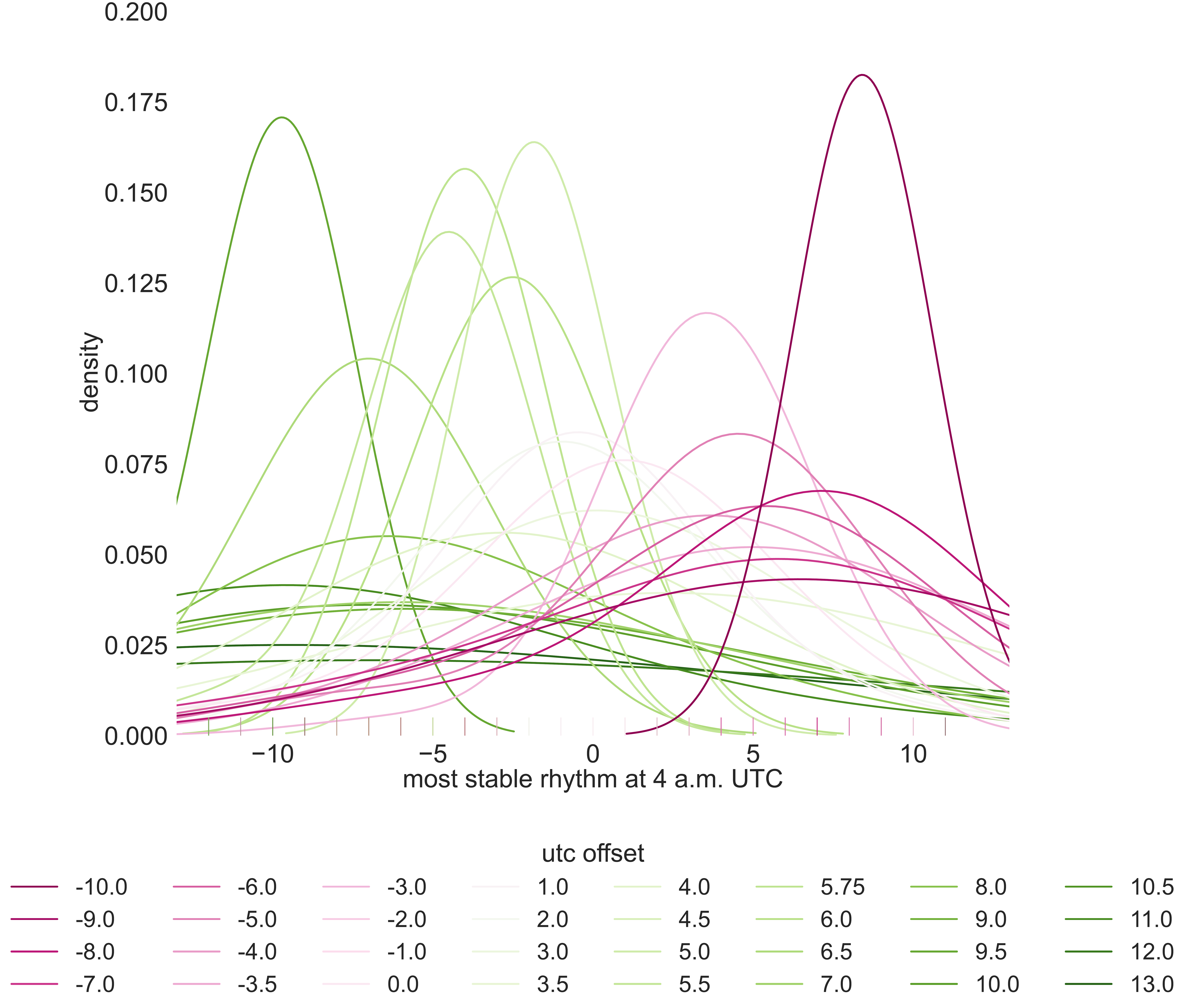}
    
  \end{minipage}
  
  \vspace{0.6cm}
  
  \caption{Frequency-domain feature distributions. (Left) The wavelet power spectrum appears noisy, in time-zone aggregates; however, analyses will show it still carries usable information. (Right) Similarly to activity lulls in the time domain, the hour of maximum rhythm stability appears to correspond consistently with 4:00 AM local time.}
  \Description{}
  \label{fig:frequency_domain_features}
\end{figure}

\paragraph{Time domain:} The left panel in Figure~\ref{fig:time_domain_features} shows the mean hourly activity distributions aggregated by UTC offset, with 95\% confidence intervals. The signal is immediately legible: the characteristic trough of the daily activity cycle---the window when the bulk of a community's users are asleep---shifts systematically across panels in lock-step with the UTC offset. Western time zones exhibit their minimum in the early UTC hours; Eastern time zones push it progressively later. Despite this pattern being more pronounced in more data-rich time zones, the shape of the cycle is otherwise broadly consistent across offsets, reflecting the shared circadian biology of human populations regardless of location.

The right panel in Figure~\ref{fig:time_domain_features} translates this observation into a scalar feature by extracting, for each subreddit, the hour of minimum activity, and plotting the Gaussian density estimation for all subreddits in each UTC offset. The distributions are well-separated and tightly peaked: the activity minimum falls consistently close to 4~a.m. local time across virtually all time zones. This regularity is the empirical foundation of our simplest inference heuristic, which requires no reference pool and no model fitting, but only the alignment of the activity lull to 4~a.m. local time.

\paragraph{Frequency domain:} The left panel in Figure~\ref{fig:frequency_domain_features} shows the mean Morlet wavelet power at the 24-hour frequency, averaged by UTC offset. Unlike the activity histograms, the power profiles do not present easily interpretable patterns: the curves are largely flat and visually similar across time zones, with no obvious shift analogous to the one observed in the time domain. This does not imply that the frequency-domain representation is uninformative: as we will show in the remainder of the results section, it carries a sizable predictive signal when used in a reference-based inference framework. This suggests that the information it encodes is of a different character than the simple phase offset visible in the histograms.

The right panel in Figure~\ref{fig:frequency_domain_features}  shows the distribution of the hour of maximum phase stability extracted from the wavelet decomposition, and aggregated similarly to the time-domain activity lull. Here, the picture resembles the time-domain result closely: the most stable phase of the 24-hour rhythm aligns near 4~a.m. local time across time zones. This convergence is reassuring: two independent representations of the same underlying signal, derived through entirely different mathematical operations, point to the same biological anchor.

Next, we will use these features as the basis for inference, comparing accuracy and computational cost systematically across the full ground-truth dataset.

\subsection{RQ1: How Reliable is Time Zone Inference Based on Time Series Data? }
\label{subsec:RQ1}
\subsubsection{Overall Performance and Model Comparison}
Our primary objective was to determine whether a community's geographic time zone could be reliably inferred from its temporal footprint. 
To evaluate this, we benchmarked our two primary paradigms across both the time and frequency domains. 

\begin{table}[h!]
  \centering
  % Rimpicciolisce proporzionalmente la tabella all'esatta larghezza della pagina
  \resizebox{\textwidth}{!}{%
  \begin{tabular}{l c c c c c}
    \toprule
    \textbf{Method} & \textbf{Accuracy} & \textbf{Weighted Kappa} & \textbf{Circular Correlation} & \textbf{Mean Circular Error} & \textbf{Weighted F1} \\
    \midrule
    Activity Counts & \textbf{0.74} & \textbf{0.92} & 0.95 & \textbf{0.48} & \textbf{0.73} \\
    Activity Counts Smooth & 0.72 & 0.91 & 0.95 & 0.50 & 0.71 \\
    Activity Lull & 0.22 & 0.85 & 0.95 & 1.03 & 0.24 \\
    Activity Lull Smooth & 0.13 & 0.85 & \textbf{0.96} & 1.16 & 0.15 \\
    Rhythm & 0.55 & 0.69 & 0.72 & 1.61 & 0.54 \\
    Most Stable Rhythm & 0.39 & 0.64 & 0.68 & 2.16 & 0.40 \\
    \midrule
    Baseline (Dummy) & 0.13 & 0.00 & 0.00 & 4.63 & 0.13 \\
    \bottomrule
  \end{tabular}%
  }
   \caption{Performance metrics across all evaluated time zone inference methods. Time-domain methods achieve better results, in particular reference-based time-domain methods.}
   \label{tab:metrics}
\end{table}

As detailed in Table \ref{tab:metrics}, the \texttt{Activity Counts} method achieved the highest overall performance across all evaluated metrics. It yielded an exact match Accuracy of 0.74, a Weighted F1-score of 0.73, and a Weighted Kappa of 0.92. Furthermore, it minimized the Mean Circular Error to just 0.48 hours. Applying cyclic smoothing to this method (\texttt{Activity Counts Smooth}) produced similarly robust results, with an Accuracy of 0.72 and an error of 0.50 hours. All substantive methods vastly outperformed the \texttt{Baseline (Dummy)}, which generated a random exact match accuracy of 0.13 and a Mean Circular Error of 4.63 hours. 

Note that the time-domain biological anchor methods successfully captured the macro-level cyclical trend but struggled with exact-hour classification. The \texttt{Activity Lull Smooth} method maintained an exceptionally high Circular Correlation of 0.96, indicating that the model perfectly identified the global directional sequence of the time zones (ordering communities correctly from West to East). However, its exact Accuracy dropped to 0.13, and its Mean Circular Error rose to 1.16 hours. The raw \texttt{Activity Lull} similarly yielded a low accuracy of 0.22.

\begin{figure*}[h!]
  \centering
  \includegraphics[width=0.9\textwidth]{
  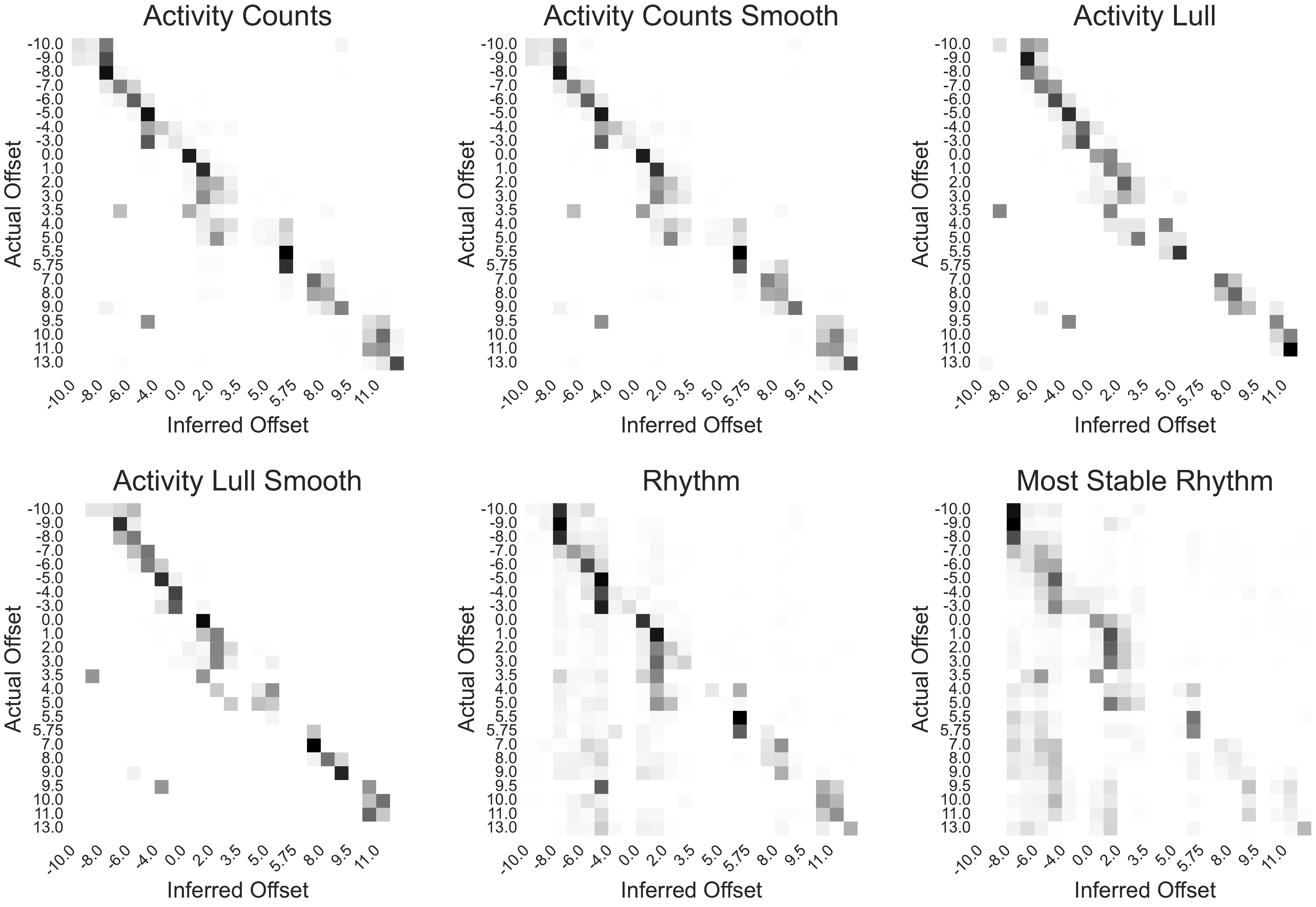}
  \vspace{0.2cm}
  \caption{Confusion matrices of actual vs. inferred UTC offsets. A strong diagonalization of the predictions is visible in the time-domain data (\texttt{'Activity'}), which is partially lost in the time-frequency domain data \texttt{'Rhythm'}. }
  \label{fig:rq1_confusion}
\end{figure*}

A visual inspection of the confusion matrices (Figure \ref{fig:rq1_confusion}) clarifies this discrepancy. The predictions for the \texttt{Activity Lull} methods remain predominantly diagonalized, demonstrating that the models do not suffer from severe or arbitrary misclassifications. Instead, the diagonal is "thickened" or spread across immediately adjacent cells. This variance is a direct result of the rigidity of the biological anchor assumption. While communities universally experience a sleep lull, fixing that lull strictly at 4:00 AM UTC fails to account for natural local variance (e.g., populations whose deepest rest naturally occurs at 3:00 AM or 5:00 AM). Consequently, the model frequently misses the exact ground-truth offset by $\pm 1$ or 2 hours. This structural bias erodes strict Accuracy while maintaining a strong overall Circular Correlation.

\subsubsection{Analysis of Error Distributions}

\begin{figure}[h!]
  \centering
  \includegraphics[width=\textwidth]{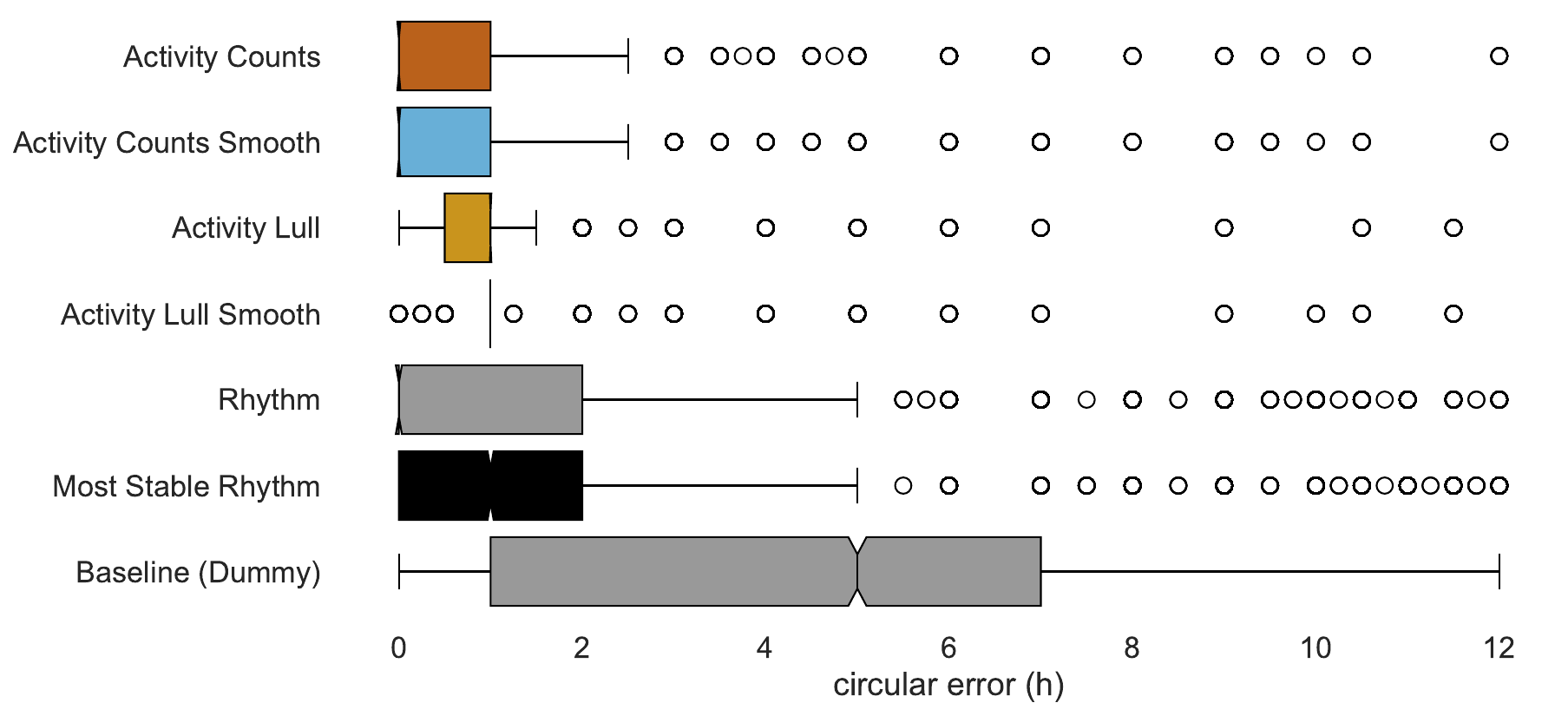}
  \caption{Distribution of absolute circular errors. The time-domain reference methods maintain tight error margins near zero. The biological lull methods exhibit a systematic median shift of 1 hour. The wavelet-based rhythm methods display extreme outlier behavior, with errors reaching up to 12 hours (anti-phase inversions).}
  \Description{A horizontal boxplot showing the distribution of circular errors in hours for each inference method. Activity counts are tightly packed near 0. Activity lull methods have their median shifted to 1. Rhythm methods show massive interquartile ranges and long tails of outliers extending to 12 hours.}
  \label{fig:circular_error}
\end{figure}

To understand the nature of the classification failures and quantify the variance across all methods, the absolute circular errors are plotted in Figure \ref{fig:circular_error}. The data reveals distinct error profiles for each methodological paradigm.

The interquartile range for the reference-based time-domain methods (\texttt{Activity Counts} and \texttt{Activity Counts Smooth}) is tightly constrained between 0 and 1 hour. These models consistently yield the highest precision with minimal deviation, confirming that matching the full 24-hour distribution successfully localizes the temporal signature.

In contrast, the biological anchor methods (\texttt{Activity Lull} and \texttt{Activity Lull Smooth}) display a systematic error shift. As shown in the horizontal boxplots, their median circular error sits exactly at 1 hour. This aligns perfectly with the visual evidence from the confusion matrices, where the predictions form a "thickened" diagonal spread across immediately adjacent cells. The model misses the exact ground-truth offset by $\pm 1$ hour, eroding strict Accuracy while still maintaining the strong overall Circular Correlation reported in Table \ref{tab:metrics}. 

Finally, the frequency-domain representations struggled significantly with extreme misclassifications. The boxplots for \texttt{Rhythm} and \texttt{Most Stable Rhythm} exhibit massive interquartile ranges and severe outlier behavior, with a long tail of circular errors stretching up to 10 and 12 hours. In a 24-hour circular space, an error of 12 hours represents a complete anti-phase inversion.

\subsubsection{Geographic Distribution of Inference Errors}
Certain regions may be more affected by inference errors than others, due, e.g., to user activity from a location being more scarce, or attracting activity from users from different locations. 
To ensure that the method is accurate across geographic regions, we map the distribution of errors for the \textit{location\_subreddits} dataset. In this section, we focus on the most accurate method, Activity Counts; however, the analysis yields qualitatively similar results for the remaining methods.

\begin{figure}[h!]
  \centering
  \includegraphics[width=\textwidth]{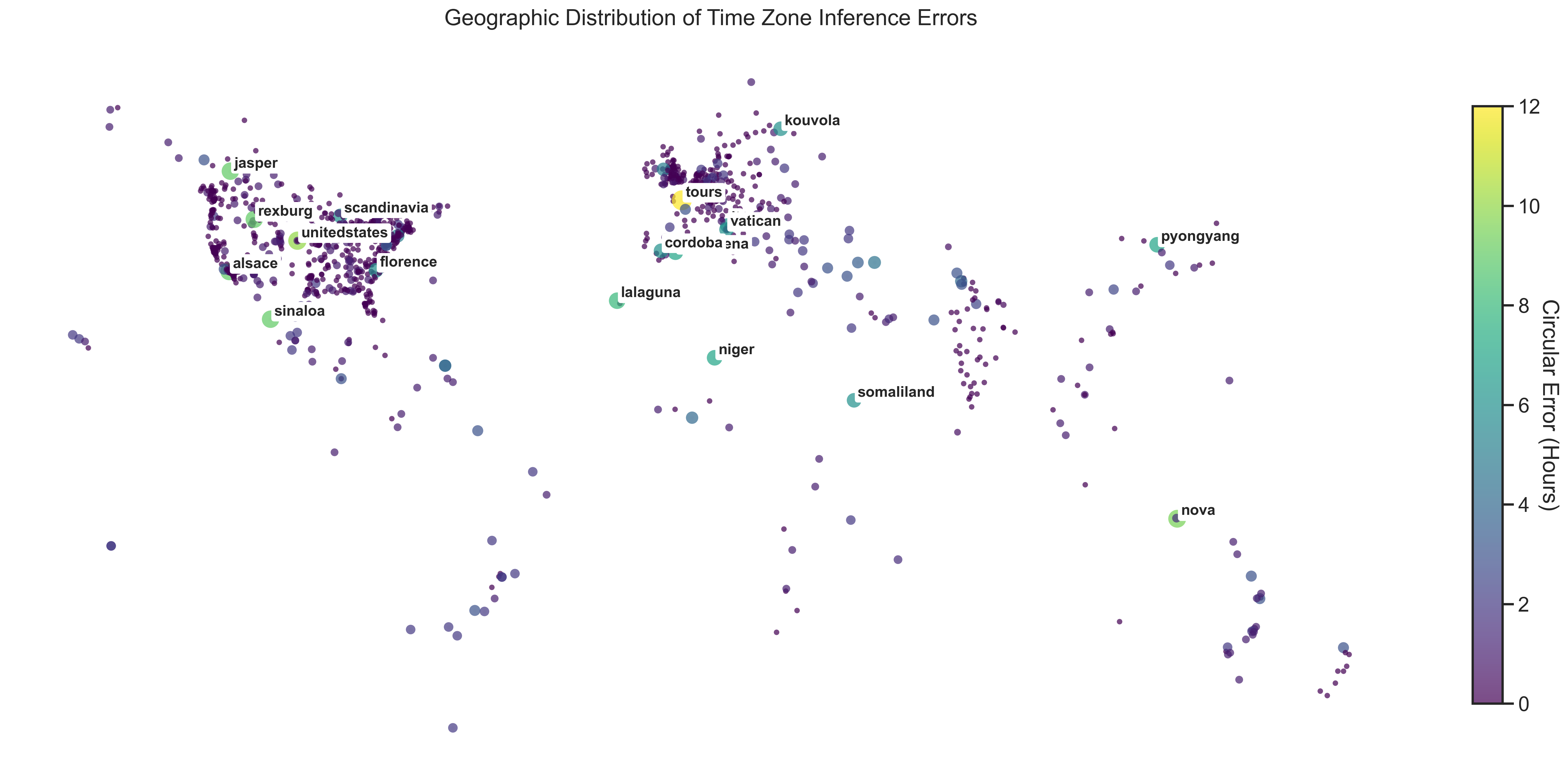}
  \caption{Geographic distribution of time zone inference errors. High-error nodes (yellow/light green) are not randomly scattered but represent specific socio-technical edge cases, such as highly international hubs (vatican), strictly moderated enclaves (pyongyang), vast multi-timezone regions (unitedstates).}
  \Description{A world map with scattered dots representing subreddits. Most dots are dark blue (low error). Scattered light green and yellow dots (high error) are annotated with names like Vatican, Pyongyang, and Scandinavia.}
  \label{fig:error_map}
\end{figure}

Figure \ref{fig:error_map} highlights the spatial concentration of inference failures. The overall accuracy of the reference-based model is high, and mapping the absolute circular errors geographically does not reveal a spatial concentration of misclassifications. Instead, it highlights distinct socio-technical edge cases that disrupt temporal inference:
\begin{enumerate}
    \item \textbf{International Tourist Hubs:} Communities such as \textit{r/vatican} and \textit{r/tours} exhibit elevated errors. Because these subreddits are dominated by international tourists and foreign observers rather than residents, their activity rhythms are completely detached from the physical solar time of the location.
    \item \textbf{Information Blackouts and Enclaves:} Nodes like \textit{r/pyongyang} show extreme anti-phase errors. Due to strict national internet controls, activity in these subreddits is generated entirely by external users (often from the US or Europe), resulting in a temporal footprint that directly contradicts their physical locations.
    \item \textbf{Geographic Broadness:} Labels spanning vast geographic areas (such as \textit{r/unitedstates}) force the model to compute an average rhythm across multiple distinct time zones, naturally introducing a $\pm 1$ or 2-hour offset error depending on where the user base is most densely concentrated.
\end{enumerate}

Finally, a few of the largest errors are actually not errors to begin with: due to the Nominatim service used to build our ground-truth, \textit{alsace}, \textit{scandinavia}, and \textit{florence} were placed similarly-named locations (i.e., Alsace, California; Scandinavia, Wisconsin; and Florence, South Carolina). The fact that our method correctly recovers the correct location for these subreddits further speaks to its robustness. 

\subsection{RQ2: How Does Inference Quality Vary with Data Availability?}
\label{subsec:RQ2}

Our second objective was to evaluate how inference quality varies with data availability. Real-world social media datasets are heavily skewed; while a few default communities generate millions of interactions, the vast majority of local or niche forums are small and sparse \cite{avalle2024persistent}. We needed to verify that our methodology was not exclusively reliant on high-volume traffic.

To test this, we evaluated how inference reliability, measured by Circular Correlation, degraded as the total volume of community comments decreased from $10^7$ down to $10^2$, and as the temporal observation window shrank from $10^3$ down to $10^1$ active days. The results of this analysis are plotted in Figure \ref{fig:data_scarcity} (see Appendix Figure \ref{fig:app_accuracy} for similar results regarding accuracy).

\begin{figure}[h!]
  \centering
  \includegraphics[width=\textwidth]{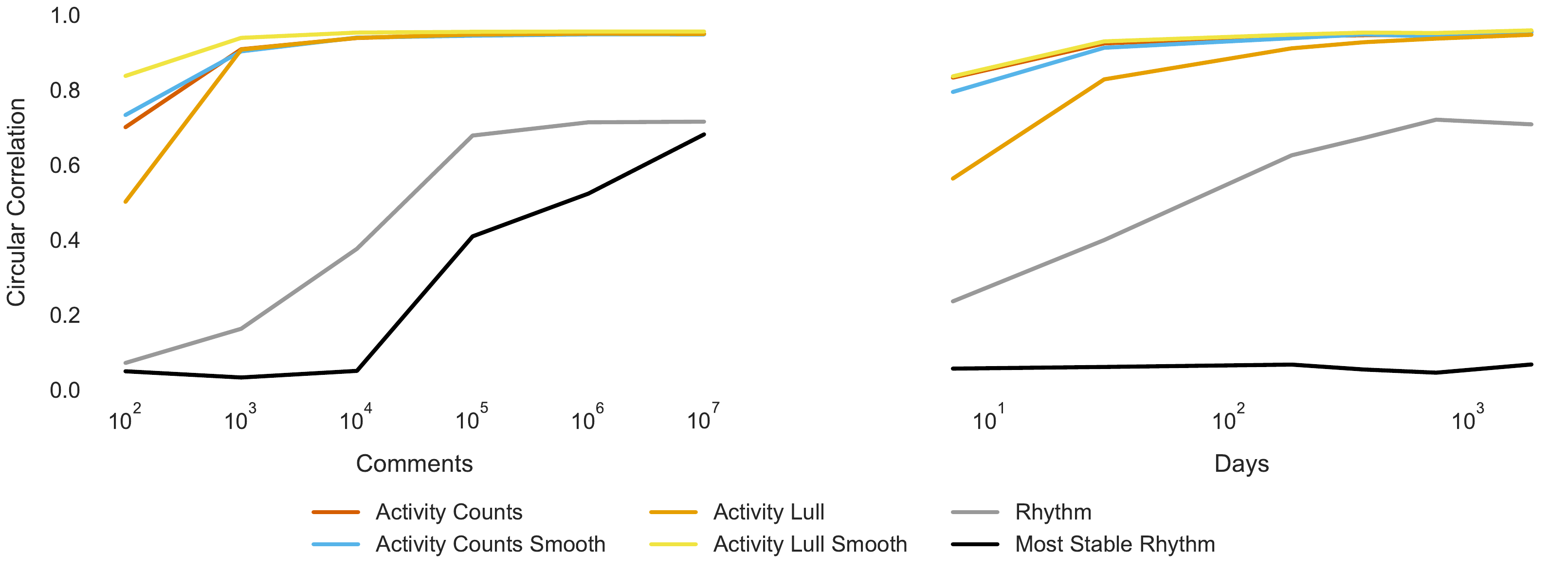}
  \caption{Inference reliability relative to data volume. Time-domain representations (Activity Counts and Activity Lull) remain remarkably robust even at low data volumes. In contrast, frequency-domain methods (Rhythm and Most Stable Rhythm) degrade rapidly under conditions of data scarcity.}
  \Description{Two line graphs plotting Circular Correlation against the number of comments and observation days. The time-domain methods stay relatively flat near the top of the y-axis, while rhythm methods drop sharply towards 0.4 at lower data volumes.}
  \label{fig:data_scarcity}
\end{figure}

\subsubsection{Robustness of Time-Domain Methods}

We found that time-domain representations remained highly robust under severe data scarcity. As shown in Figure \ref{fig:data_scarcity}, even at the lowest evaluated data volumes ($10^2 $ comments or 10 days), the Circular Correlation of the \texttt{Activity Counts} and \texttt{Activity Lull} models remained exceptionally stable between 0.7 and 0.9. We attribute this stability to the fact that the simple mathematical normalization of a 24-hour histogram proves highly resilient to sparsity. It effectively aggregates sparse, intermittent signals across weeks or months into a recognizable, stable circadian shape, requiring surprisingly few actual data points to establish a baseline routine.

\subsubsection{Degradation of Frequency-Domain Methods}

Conversely, the frequency-domain models degraded significantly under sparse conditions. As the number of comments dropped below $10^5$, the correlation of the \texttt{Most Stable Rhythm} and standard \texttt{Rhythm} methods plummeted towards and then below 0.4. Notably, the \texttt{Most Stable Rhythm} approach requires years of data before it reaches usable performance levels. This shows that wavelet-based approaches, although they capture complementary information to time-domain methods, may not be suitable in data-scarce scenarios.

\subsection{RQ3: What is Reddit's Geographical Distribution, and How did it Evolve?}
\label{subsec:RQ3}

The previous sections established the reliability of a parsimonious heuristic, Activity Lull, on a controlled, geographically labeled dataset. As a final step, we apply this model at scale across the broader, unlabeled Reddit ecosystem to show its practical utility in the wild. 

This large-scale evaluation is structured into two distinct phases: first, we focus on specific topical categories to qualitatively validate cultural and geographic clustering of subreddits; then, the model is deployed across the entire historical dataset to track how the platform's global footprint has shifted from 2005 to 2024.

\subsubsection{Qualitative Validation of Semantic Clusters}

% Communities centered around specific topics should naturally cluster around the geographic regions where those topics are most culturally relevant. Take, for example, a sport like cricket: you would expect communities that talk about cricket to be based predominantly in India or in its proximity. Analogously, you would expect communities focused on or related to Hinduism to concentrate between UTC+4 and UTC+6. Similarly, to validate the model's performance on non-geotagged data, the inferred UTC offsets were mapped against specific semantic categories.

% Figure \ref{fig:semantic_clusters} illustrates this alignment using four distinct thematic categories: (a) Place; (b) Business, Economics, and Finance; (c) Food and Drink; and (d) Sports.

\begin{figure}[h!]
    \centering
    \includegraphics[width=\textwidth]{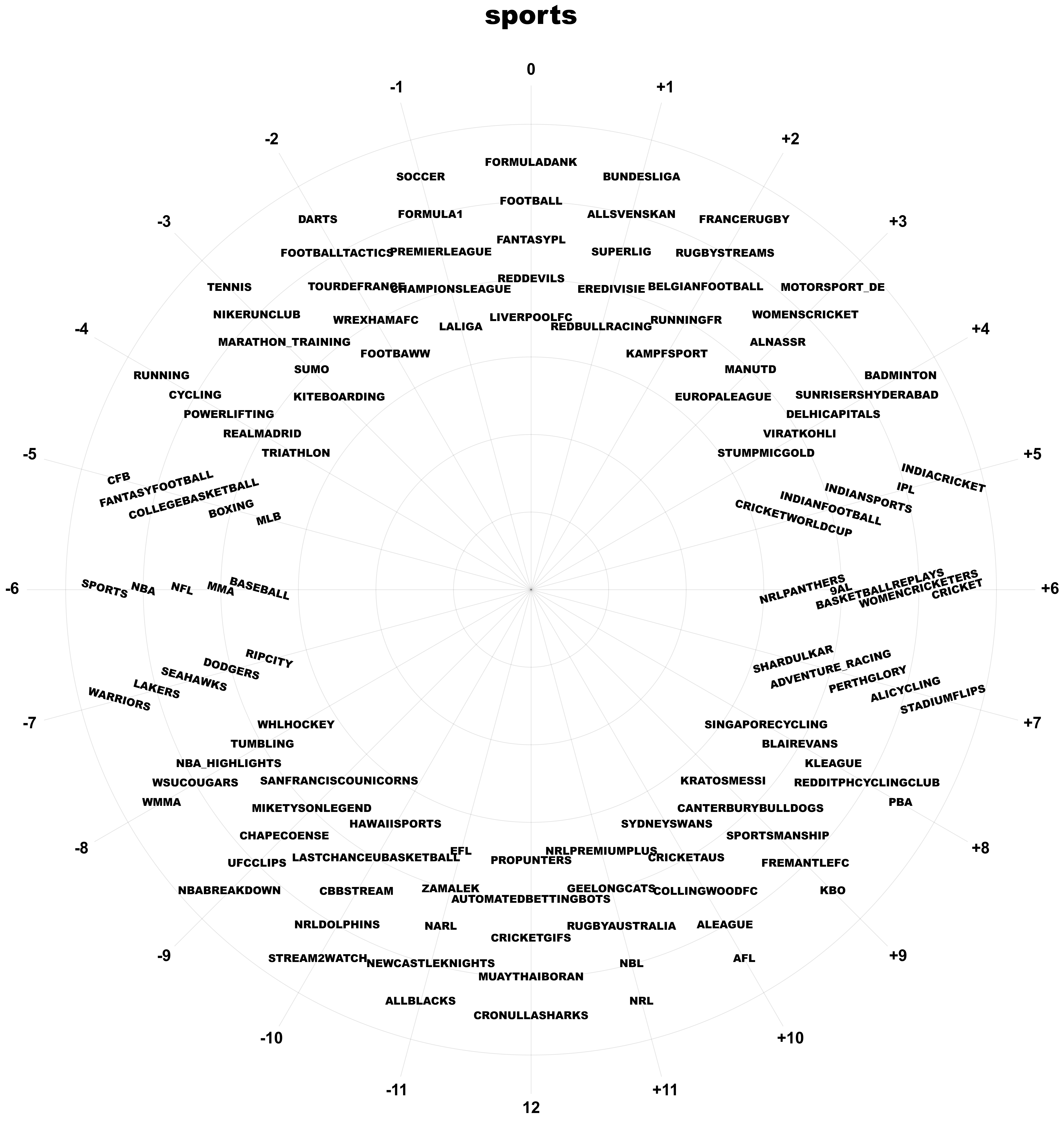}
    \caption{Inferred UTC offsets for subreddits in the Sports category.}
    \label{fig:sports}
\end{figure}

Beyond quantitative accuracy metrics, a compelling test of the method is whether the inferred time zones reflect recognizable cultural geography --- whether the communities placed at a given offset are the communities one would expect to find there. Figure~\ref{fig:sports}, and Figures~\ref{fig:semantic_clusters},\ref{fig:subreddit_geo_map_business_economics_and_finance_angled}, and \ref{fig:subreddit_geo_map_food_and_drink_angled} in the appendix, display the inferred time zones of subreddits across four thematically distinct categories: sports, business and finance, food and drink, and place. Each radial plot positions subreddits at their inferred UTC offset. 

Across all four categories, the method recovers a legible map of the world from nothing but posting timestamps. The place category (Figure~\ref{fig:semantic_clusters} in appendix) offers a direct sanity check: even after removing the subreddits used as anchors, subreddit names in this category are often explicit geographic references. The method correctly places, for example,  \texttt{Europe} near UTC$0$; \texttt{Disneyland} at UTC$-$7; \texttt{IndiaSocial} at UTC$+$5 to UTC$+$5.5; \texttt{ChinaIRL} at UTC$+$7; and \texttt{Australia} at UTC$+$10. Errors, when they occur, are typically small and geographically plausible: a community displaced by one hour to a neighboring offset, rather than placed on the wrong continent. 

Similar results apply to the business and finance category (Figure~\ref{fig:subreddit_geo_map_business_economics_and_finance_angled} in appendix): communities with explicit national markers, such as \texttt{WallStreetBets}, \texttt{IndianStockMarket}, \texttt{AusFinance}, \texttt{NZBusiness}, \texttt{MalaysianPF}, are correctly placed at their respective national offsets. The food and drink chart (Figure~\ref{fig:subreddit_geo_map_food_and_drink_angled} in appendix) reinforces the same pattern: for example, \texttt{HeelHollandKookt} and \texttt{Geldzaken} appear near UTC$+$1$-$2 (Dutch content), \texttt{MumbaiFoodPhotos} and \texttt{IndiaFood} near UTC$+$5$-$6, and Australian food communities near UTC$+$8$-$10. 

The sports chart provides perhaps the clearest validation (Figure~\ref{fig:sports}). The left half of the dial (negative offsets) is dominated by North American content: \texttt{NFL}, \texttt{NBA}, \texttt{MLB}, \texttt{CFB}, and city-specific franchises such as \texttt{Lakers}, \texttt{Warriors}, and \texttt{Seahawks} cluster tightly between UTC$-$5 and UTC$-$8, consistent with U.S.\ time zones. Moving toward zero, European football communities appear in their expected positions: \texttt{PremierLeague}, \texttt{ChampionsLeague}, \texttt{Bundesliga}, and \texttt{Eredivisie} sit near UTC$-$1 to UTC$+$2, while \texttt{Superlig} and \texttt{FranceRugby} align further east. On the opposite side of the dial, Australian rules football and cricket communities --- \texttt{AFL}, \texttt{NRL}, \texttt{CricketAus}, \texttt{ALeague} --- occupy the UTC$+$8 to UTC$+$10 range, and Indian cricket communities (\texttt{IPL}, \texttt{IndiaCricket}) appear near UTC$+$5.5, matching India Standard Time.

Although largely aligning with expected geographical distributions, these case studies reveal a limitation that is instructive rather than surprising. Globally oriented communities --- such as, \texttt{Investing} or \textit{Cooking} --- cluster in U.S. time zones despite having internationally distributed user bases. This is an expected failure mode: when a community's membership and activity span multiple continents without a dominant local majority, the inferred time zone reflects the plurality rather than a unique geographic center. Reddit's historical U.S.-centric user base means that generalist communities tend to be pulled toward North American offsets regardless of their nominal scope. Taken together, these four charts complement the quantitative evaluation: our method recovers a coherent, interpretable geography of online culture.

\subsubsection{Measuring Reddit's Globalization}

Is Reddit a truly global social network? Has its subreddit base changed at all from its birth in 2005 to now?

Having validated the method's scalability, we apply the inference pipeline to \texttt{all\_subreddits} to track macro-level temporal shifts and observe the platform's evolution from 2006 to 2024.

\begin{figure}[h!]
    \centering
    \includegraphics[width=\textwidth]{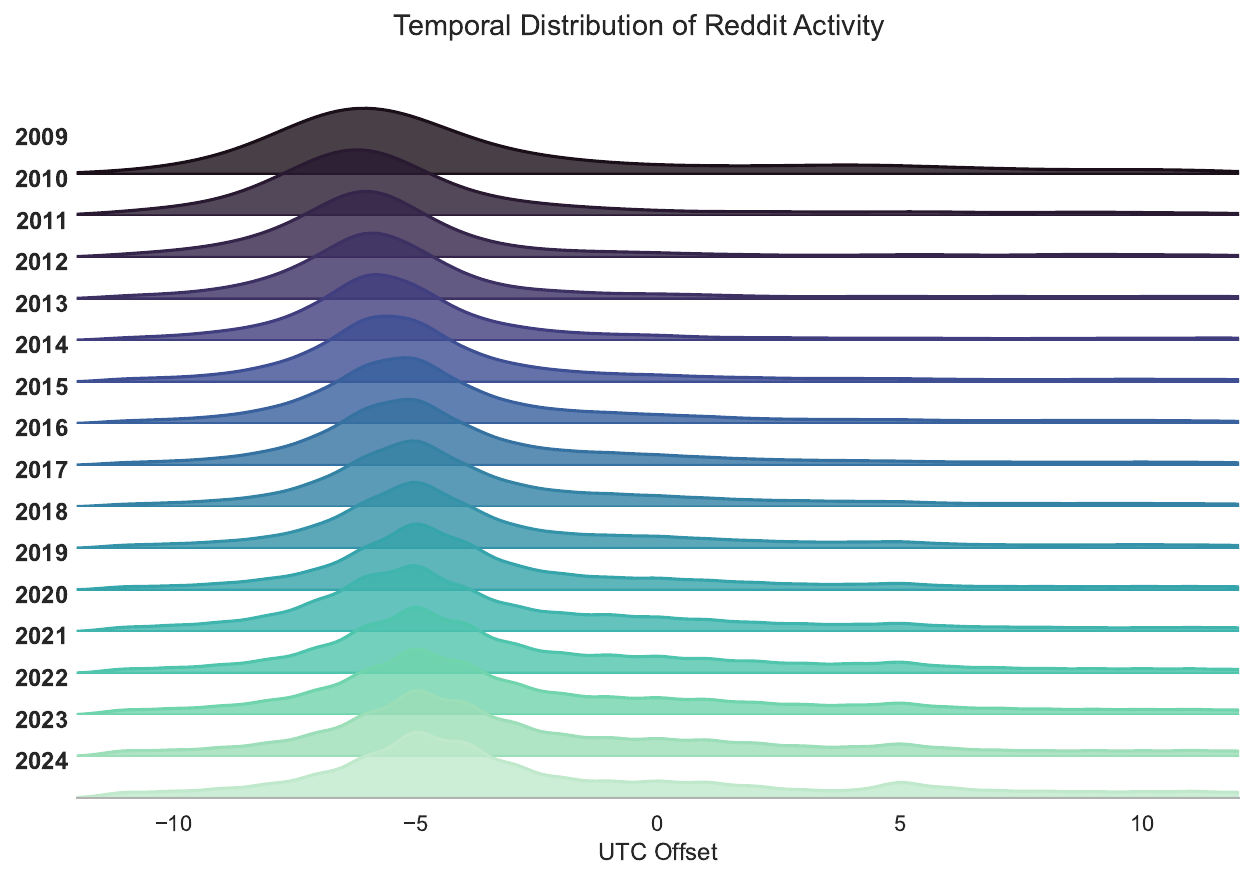}
    \caption{Longitudinal distribution of Reddit activity (2009--2024). The density curves demonstrate a historical progression from a nearly exclusive US-centric base (UTC-5 to -8) towards a rapidly growing European and Asian presence. }
    \label{fig:evolution}
\end{figure}

As shown in Figure \ref{fig:evolution}, the temporal density of the platform has undergone a fundamental structural shift. In its early years (2009--2014),\footnote{We omit years 2005--2008 because the poor Gaussian fits on the scarce data in these initial years introduce artifacts in the visualization, making it less interpretable.} the platform's temporal footprint was almost exclusively anchored in North America (UTC -5 to -8). Over the following decade, the distribution visibly flattens and spreads eastward. By 2024, distinct secondary and tertiary peaks have solidified around UTC 0 (Western Europe) and positive offsets, in particular +5 (India), indicating the platform's transition into a globalized ecosystem.

\begin{figure}[h!]
    \centering
    \includegraphics[width=\textwidth]{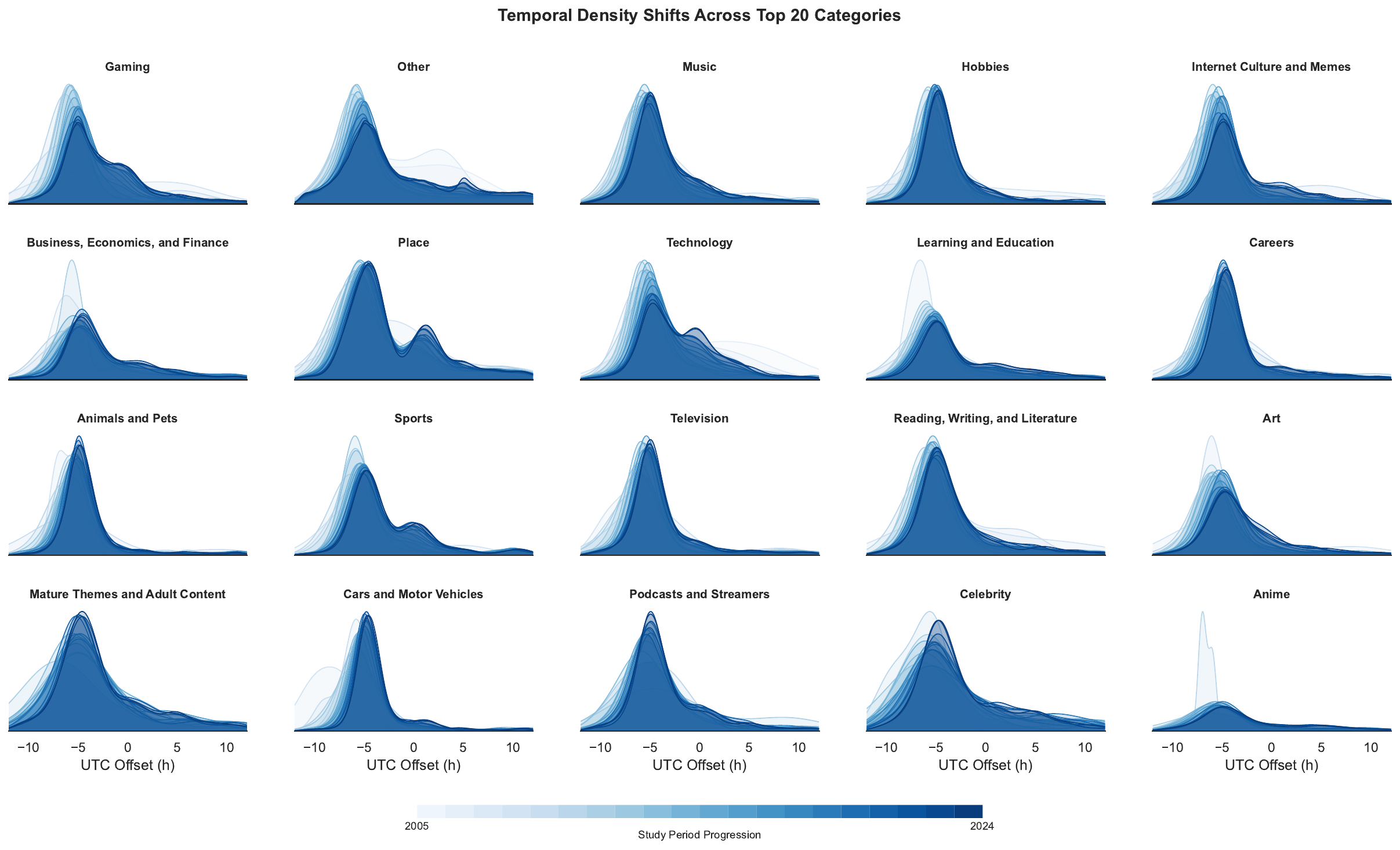}
    \caption{Longitudinal distribution of Reddit activity per subreddit topic. Topics like Technology, Place, and Sports show a distinctive bimodal distribution whereby the peaks align with the US and Europe time zones, with the European peak becoming more prominent in more recent years. Other topics show unimodal distributions, consistently dominated by a US-centric peak but showing longer tails---and thus, global audiences---like Anime, Learning, and Art.}
    \label{fig:topic_evolution}
\end{figure}

This 'globalization', however, is not uniform across all topics. Figure \ref{fig:topic_evolution} disaggregates this temporal density by the top 20 semantic categories. The granular category breakdown reveals distinct cultural diffusion rates. Digital-native and entertainment categories such as \textit{Gaming}, \textit{Place}, and \textit{Anime} exhibit the most aggressive temporal spread, showing rapid and continuous adoption across European and Asian time zones. Among these shifting topics, the \textit{Technology} category presents a particularly striking evolutionary pattern. Over the 20-year observation period, it developed the most notable and structurally distinct secondary peak exactly at UTC 0. This emergent distribution, resembling the bimodal distribution of \textit{Place}, indicates a sustained adoption by European bases, effectively splitting the platform's global technical discourse into two almost co-dominant temporal hubs (North America and Western Europe).

Conversely, categories tied to regional broadcasting schedules, localized job markets, and established domestic routines resist this globalizing trend. Topics such as \textit{Television}, \textit{Careers}, \textit{Animals and Pets}, and \textit{Reading, Writing, and Literature} retain much sharper, rigid distributions. Despite the platform's overall eastern expansion, these communities remain heavily anchored to their historical US-centric time zones, primarily between UTC -5 and -8. This divergence confirms that while digital and internet-native communities fluidly cross geographic boundaries, topics governed by regional media releases, standard corporate working hours, and localized offline hobbies remain strictly bound to their legacy geographic anchors.

While the density distributions in Figure \ref{fig:evolution} and \ref{fig:topic_evolution} illustrate the qualitative shift in subreddit activity, we can formalize this geographic decentralization by quantifying the structural evolution of the communities themselves. To measure the extent to which the creation of new subreddits is dispersed across the globe, we calculate the Gini Coefficient of the subreddit temporal distribution for each year. In this context, the Gini index measures the concentration of communities across the 24 inferred time zones: a value approaching 1 indicates total concentration (all subreddits anchored to a single time zone), while a value approaching 0 indicates perfect dispersion across the 24-hour cycle.

\begin{figure}[h!]
    \centering
    \includegraphics[width=\textwidth]{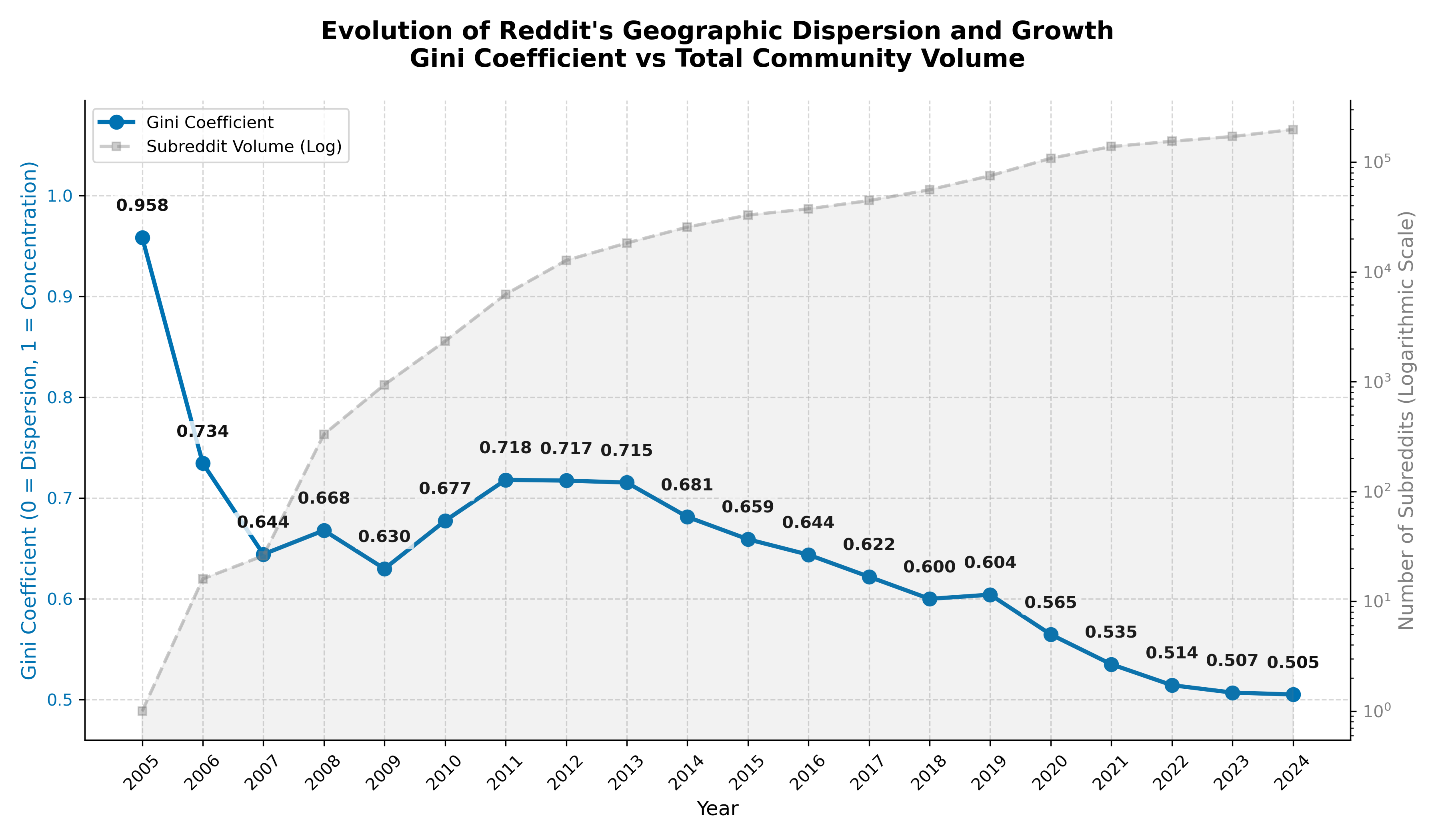}
    \caption{Evolution of Reddit's Geographic Dispersion (2005--2024). The Gini Coefficient, calculated on the distribution of subreddits across inferred time zones, shows a steady decline from highly concentrated ($0.718$ in 2011) to increasingly decentralized ($0.505$ in 2024).}
    \label{fig:gini_evolution}
\end{figure}

As depicted in Figure \ref{fig:gini_evolution}, the platform's history since 2011 can be characterized by a clear quasi-monotonic trend toward decentralization, with the slight exception of 2019. During its early consolidation phase (2011--2013), the Gini index stabilized at a highly concentrated value of roughly $0.71$, confirming that the rapid initial creation of subreddits was tightly bound to the dominant North American time zones. However, from 2014 onward, the index experiences a steady and practically uninterrupted decline, reaching $0.505$ by 2024. This significant drop confirms that the expansion of Reddit's infrastructure is no longer monopolized by its founding geographic centre; instead, communities are increasingly distributed across the global temporal spectrum.

To understand the specific geographic drivers behind this falling Gini coefficient, we computed a relative Growth Index for each inferred time zone. To symmetrically capture both the exponential emergence of new communities and the relative historical absence of others, we apply a $\log_2$ fold change transformation relative to a base year (2012). In this logarithmic space, the 2012 baseline is normalized to zero; positive values denote exponential community growth relative to the baseline, while negative values indicate a fractional density.

\begin{figure}[h!]
    \centering
    \includegraphics[width=\textwidth]{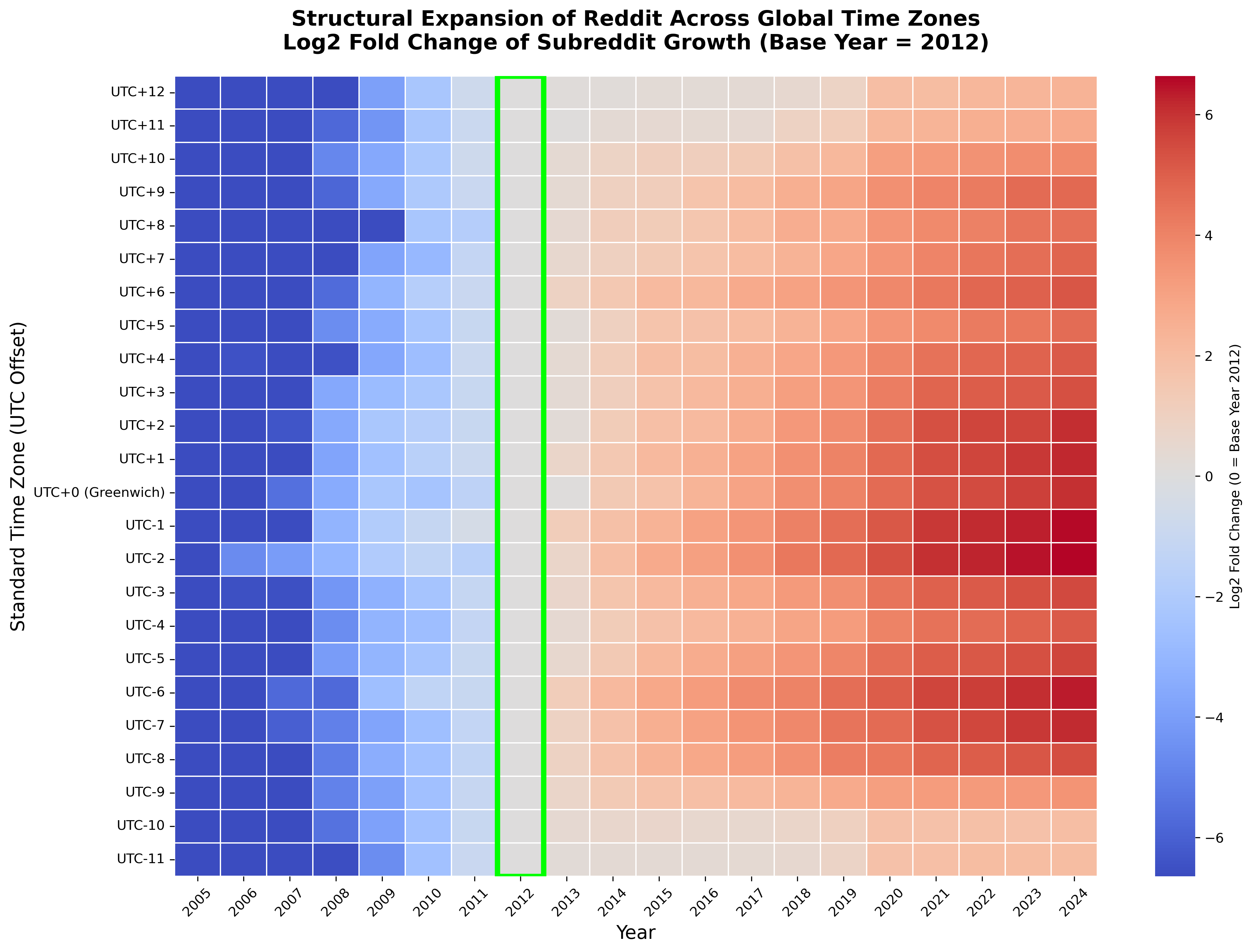}
    \caption{Heatmap of the Log2 Fold Change in subreddit volume (Base Year = 2012). Following the 2011--2013 consolidation phase, the platform experienced global expansion. While the established North American core (UTC -5 to -8) maintains strong, sustained growth, a broader array of European, Atlantic, and Asian time zones exhibit comparable or slightly higher relative growth rates, actively broadening the platform's temporal footprint.}    \label{fig:growth_heatmap}
\end{figure}

Indeed, the heatmap in Figure \ref{fig:growth_heatmap} elegantly visualizes the mechanics of Reddit's globalization by isolating the relative velocity of community formation. Moving forward from the 2012 baseline, the platform does not experience a shift of activity away from its origins, but rather a simultaneous, universal expansion, as the historically dominant North American core (UTC -5 to -8) continues to grow robustly. Crucially, however, this North American expansion is matched—and marginally exceeded—by the activation of a much wider array of global time zones. A contiguous block spanning from UTC +2 through UTC -2 exhibits intense compounding growth, reaching peak log-fold changes between 5 and 6. Concurrently, Asian time zones (such as UTC +8 and +9) experience an increase in relative community formation.

This specific growth dynamic—where a larger number of historically peripheral time zones grow at rates comparable to, or slightly higher than, the established core—perfectly explains the falling Gini coefficient observed in Figure \ref{fig:gini_evolution}. The geographic decentralization of Reddit is not caused by the stagnation of its US-centric user base. Instead, it seems to be driven by the broadening of the platform's infrastructural base.

To definitively validate whether this inferred structural expansion reflects the actual geographic footprint of human users, we correlated our 2024 subreddit density data with independent, real-world internet traffic demographics. Physical user traffic estimates per country were sourced from the World Population Review (2024)\footnote{\url{https://worldpopulationreview.com/country-rankings/reddit-users-by-country}}. Correlating sovereign nations to 24-hour inferred time zones presents a mismatch, as there exists no clear one-to-one correspondence. To resolve this discrepancy, especially for transcontinental nations (e.g., the United States, Canada, Australia, and Russia), physical user allocations were proportionally disaggregated into UTC offsets. These allocation weights were derived directly from recent official state- and province-level census data (U.S. Census Bureau 2020\footnote{\url{https://www.census.gov/programs-surveys/decennial-census/decade/2020/2020-census-results.html}}, Statistics Canada 2021\footnote{\url{https://www12.statcan.gc.ca/census-recensement/2021/dp-pd/prof/index.cfm?Lang=E}}, ABS 2021\footnote{\url{https://www.abs.gov.au/census}}, INEGI 2020\footnote{\url{https://en.www.inegi.org.mx/programas/ccpv/2020/}}), ensuring that sparse regions within large nations did not disproportionately skew the baseline. Single-timezone nations were assigned 100\% weight to their standard offset. A direct linear correlation (Pearson) between the inferred peak-hour of our subreddits and the real-world physical user distribution yielded a moderate but significant coefficient of $r = 0.589$ (Spearman $\rho = 0.444$). \footnote{If instead of assuming that all of the subreddits' population is located at the inferred time zone, and instead allow it to be redistributed around this center of mass, the correlation becomes Pearson coefficient of $r = 0.895$ (Spearman $\rho = 0.662$) --- see Appendix \ref{app:deconvolution}.} This shows how the proposed models reconstruct genuine signals about the platform's geographical distribution.

\section{Limitations}
% we don't engage in the ecological fallacy: inferring on individuals based on aggregates. others do. however, inference on users would be ultimately interesting and useful.

% one time zone per subreddit, rather than a mixture

% exclude subreddits and regions with multiple time zones

% time zones rather than locations. we find a consistent signal with parsimonious methods, but these are arguably insufficient to recover finer-grained distinctions such as national borders between Germany and Austria. Other complementary information may be necessary, like dialectal variation.
Several limitations of the present work point to directions for future research, starting from how our method operates on aggregated community activity and makes no claims about the location of individual users. This is a deliberate design choice: by targeting communities rather than users, we avoid the ecological fallacy of inferring individual-level attributes from group-level patterns---a conflation that is common in the geolocation literature but for which further validation would be necessary~\cite{balsamo2019firsthand-20e,bozarth2023role-092,waller2021quantifying-ed5}. % The tradeoff is that user-level inference, which would be ultimately more useful % for many downstream applications, remains out of reach with the present approach. 
Extending timestamp-only inference to individual users is a natural next step, although it would require substantially more data per user and a careful treatment of the increased heterogeneity in individual posting schedules.
% \paragraph{One time zone per community.}
We assume that each subreddit is associated with a single time zone, treating geographic location as a fixed community attribute. This is a reasonable approximation for tightly localized communities, such as a subreddit dedicated to a specific city or region. However, this assumption becomes increasingly strained for larger or more diffuse communities, whose membership spans multiple time zones. In such cases, the observed activity pattern reflects a mixture of local rhythms, and the inferred time zone should be interpreted as a center of gravity rather than a precise location. Future work could relax this assumption by modeling communities as mixtures of time zone components, estimating both the number of geographic clusters and their respective offsets from the aggregate time series.
% \paragraph{Exclusion of multi-timezone regions.}
 As a consequence of the single-timezone assumption, we exclude from our evaluation subreddits whose geographic bounding boxes span multiple time zones. This filter disproportionately affects large countries such as the United States, Russia, and Brazil, as well as communities organized around continental or global topics. This exclusion ensures the validity of our ground truth but limits the coverage of our evaluation and the direct applicability of the method to communities organized at the national scale in geographically large countries.

% \paragraph{Time zones rather than locations.}
Perhaps most fundamentally, the methods in this analysis recover time zones rather than geographic locations. A time zone offset is a coarse spatial unit that conflates many distinct places: UTC+1, for instance, spans much of Europe, encompassing countries like Denmark, Italy, and the Czech Republic, as well as countries in the African continent like Algeria, Cameroon, and Tunisia: i.e., countries with very different languages, cultures, and institutions. Our results demonstrate that the temporal signal is strong and consistent enough to resolve offset-level distinctions reliably, but it is arguably insufficient, on its own, to recover finer-grained geographic distinctions such as the national border between Germany and Austria, or between Argentina and Chile. Resolving such distinctions would require complementary signals that carry country- or region-specific information---dialectal variation in text being a natural candidate~\cite{eisenstein2018identifying-889}---and the combination of temporal and linguistic signals in a unified inference framework represents a promising direction for future work.

\section{Conclusions}

This paper has demonstrated that the geographic time zone of an online community can be reliably inferred from nothing more than hourly activity counts. Across a broad evaluation on Reddit, the best-performing method --- comparing a community's 24-hour activity histogram against a reference pool of geolocated subreddits --- achieves a mean circular error of under 30 minutes, with an exact-hour accuracy of 74\%. A simple heuristic that anchors the activity minimum to 4 a.m. local time, requiring no supervision, recovers the correct time zone within a one-hour margin on average, nearly matching the reference-based methods at a fraction of the computational cost. Both methods apply to the long tail of small communities that dominate platforms like Reddit, being robust to both data scarcity and limited temporal observation. Applying the methods to unlabeled data reinforces these results: communities across categories ranging from sports to food to finance show a legible cultural geography of the platform, and a significant correlation between inferred community time zones and independent estimates of real user populations confirms that the temporal signal tracks actual human geography rather than mere posting patterns. We demonstrated the scalability of these methods by applying them to the entirety of Reddit: the method reveals that the platform has undergone substantial and ongoing geographic decentralization, despite a still majority U.S.-centric user base.

Taken together, these results establish community-level temporal activity as a practical, portable baseline for geographic inference in online platforms. It is resistant to individual manipulation and degrades gracefully under data scarcity. As a standalone signal, it is already informative; as a component of richer geolocation pipelines, it provides a foundation that is difficult to replicate with content-based approaches alone.

\section{Ethical Considerations}
% we comply with FAIR principles
% a number of threats posed by user geolocation/attribute inference. however, we only take aggregates that don't directly expose sensitive information about users.
All data used in this study were collected from publicly accessible sources under terms consistent with academic research use. Notwithstanding, we acknowledge that geolocation and attribute inference from social media data raise legitimate privacy concerns. Methods that recover the location of individual users from their digital traces can expose sensitive information without consent, enable surveillance, and cause harm to vulnerable populations. These risks have been extensively discussed in the literature~\cite{fiesler2018participant-fff}. Our approach is designed to sidestep these concerns by construction. We operate exclusively on aggregated community-level activity counts: the input to our method is the hourly distribution of posting volume across a subreddit, and no individual user's data is ever isolated, stored, or analyzed. The temporal signal we exploit is a population-level property of the community, not a trace attributable to any particular person. As a result, our method does not expose sensitive information about individual users, and the outputs it produces, i.e., time zone estimates for communities, carry no greater privacy risk than the public metadata already associated with those communities. Still, we recognize that even aggregate geographic inference is not without implications. Community-level location estimates could, in principle, be used to target communities based on their geography, or to make inferences about the demographic composition of a platform that individual communities might not wish to disclose. We release our method in the spirit of transparency and reproducibility, and encourage researchers who build on this work to consider these downstream uses carefully.

% %%
% %% The acknowledgments section is defined using the "acks" environment
% %% (and NOT an unnumbered section). This ensures the proper
% %% identification of the section in the article metadata, and the
% %% consistent spelling of the heading.
% \begin{acks}
% \end{acks}

\bibliographystyle{ACM-Reference-Format}
\bibliography{sample-base}

\appendix

\section{Population Location Deconvolution}\label{app:deconvolution}

Assigning a subreddit's entire user base exclusively to its peak hour ignores the longitudinal span of digital communities (e.g., a New York-centric subreddit actively engages users in California and Western Europe). 

To account for this, we applied an empirical deconvolution algorithm, framing the inferred peak hour not as a rigid boundary, but as the geographic centre of mass of a dispersed community. Using a Sequential Least Squares Programming (SLSQP) optimizer, we extracted the empirical spatial distribution that maximized the alignment between our inferred data and real-world traffic. Crucially, the optimization was subject to a strict bounding constraint ensuring non-negative percentage distributions ($0 \le w_i \le 1$) that sum to unity ($\sum w_i = 1$). Furthermore, because the 24-hour UTC cycle is inherently circular, we enforced a zero circular center-of-mass equality constraint ($\sum w_i \sin(\theta_i) = 0$, where $\theta_i = \text{shift}_i \times \frac{2\pi}{24}$). This ensured that the optimizer was allowed to distribute users into adjacent time zones, provided the mathematical directional barycenter of the subreddit remained anchored exactly to its originally inferred UTC peak.

By applying this natural demographic distribution to our inferred data, the results of which are displayed in Table \ref{tab:optimization_results}, real-world physical users correlate with our estimates with a Pearson coefficient of $r = 0.895$ (Spearman $\rho = 0.662$). This strong structural alignment confirms that the rhythmic timestamps embedded within Reddit's infrastructure contain a highly coherent and extractable footprint of the platform's global demographic.

\begin{table}[b]
    \centering
    \small
    \begin{tabular}{lrrrr}
        \toprule
        \textbf{UTC Offset} & \textbf{Real Users (\%)} & \textbf{Inferred Pure (\%)} & \textbf{Inferred Optimized (\%)} & \textbf{$\Delta$ (Opt $-$ Pure)} \\
        \midrule
        -11 &  0.00\% &  0.10\% &  1.00\% & +0.90\% \\
        -10 &  0.21\% &  0.08\% &  1.02\% & +0.94\% \\
         -9 &  0.21\% &  0.15\% &  1.72\% & +1.57\% \\
         -8 &  7.48\% &  0.30\% &  6.81\% & +6.51\% \\
         -7 &  3.85\% &  1.21\% &  5.43\% & +4.22\% \\
         -6 & 15.24\% &  5.84\% & 16.07\% & +10.22\% \\
         -5 & 22.86\% & 45.42\% & 17.48\% & -27.95\% \\
         -4 &  0.72\% & 24.28\% &  7.62\% & -16.66\% \\
         -3 &  4.73\% &  6.09\% &  2.60\% & -3.49\% \\
         -2 &  0.00\% &  2.71\% &  2.26\% & -0.45\% \\
         -1 &  0.00\% &  4.04\% &  2.55\% & -1.49\% \\
          0 &  8.27\% &  2.87\% &  5.61\% & +2.74\% \\
         +1 & 16.85\% &  2.03\% & 10.27\% & +8.24\% \\
         +2 &  2.28\% &  1.21\% &  5.08\% & +3.87\% \\
         +3 &  2.55\% &  0.54\% &  1.57\% & +1.02\% \\
         +4 &  0.34\% &  0.57\% &  1.25\% & +0.68\% \\
         +5 &  5.92\% &  0.97\% &  1.47\% & +0.50\% \\
         +6 &  0.21\% &  0.31\% &  0.97\% & +0.66\% \\
         +7 &  0.64\% &  0.28\% &  0.90\% & +0.62\% \\
         +8 &  3.76\% &  0.41\% &  2.23\% & +1.82\% \\
         +9 &  0.83\% &  0.15\% &  1.37\% & +1.22\% \\
        +10 &  2.32\% &  0.13\% &  1.91\% & +1.78\% \\
        +11 &  0.00\% &  0.22\% &  1.25\% & +1.03\% \\
        +12 &  0.74\% &  0.08\% &  1.60\% & +1.52\% \\
        \bottomrule
    \end{tabular}
    \caption{Detailed breakdown of geographic distributions across the 24-hour UTC cycle. The table compares real-world user metrics against the initially inferred density (Pure) and the deconvolved density (Optimized), highlighting the magnitude of the empirical spatial shifts ($\Delta$).}
    \label{tab:optimization_results}
\end{table}

\section{Inference Reliability Supplement}\label{app:reliability}

\begin{figure}[h!]
    \centering
    \includegraphics[width=\textwidth]{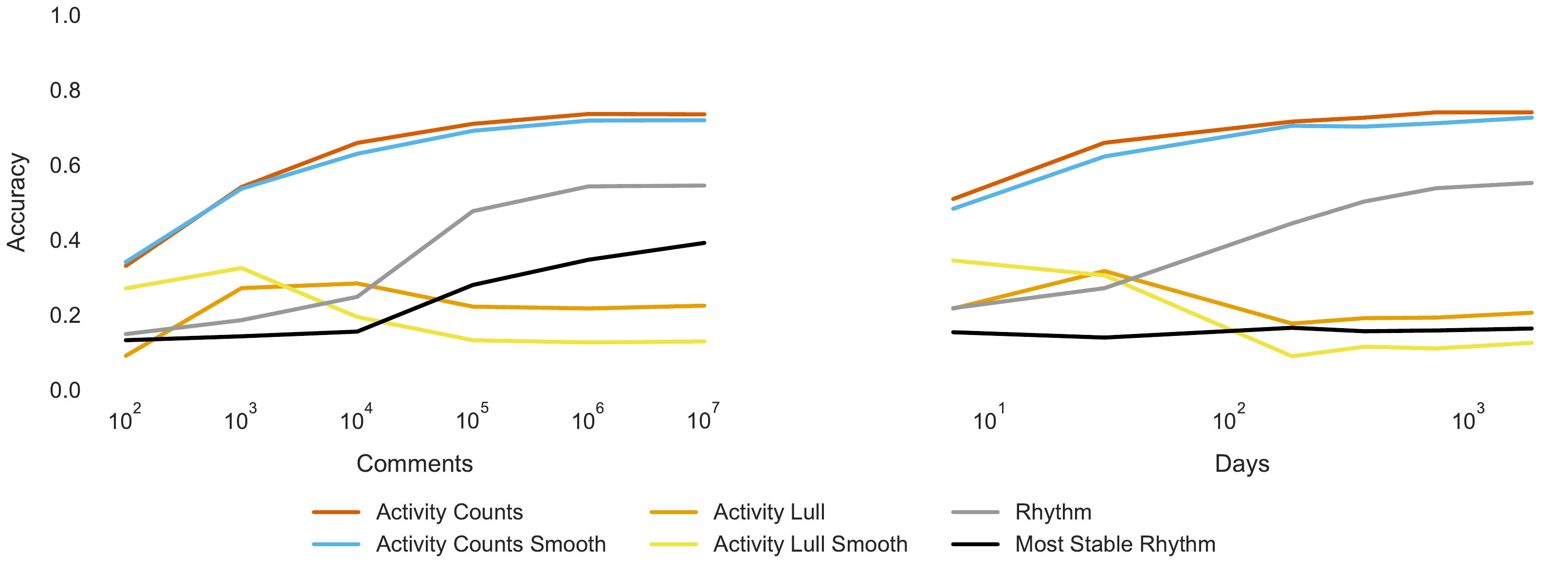}
    \caption{Degradation of inference accuracy in relation to data scarcity. The left panel illustrates accuracy as a function of the total volume of comments, ranging from $10^7$ down to $10^2$. The right panel displays accuracy relative to the temporal observation window, spanning from $10^3$ down to $10^1$ active days.}
    \label{fig:app_accuracy}
\end{figure}

\section{Time Zones for Selected Subreddit Topics}
\begin{figure}[h!]
  \centering
    \centering
    \includegraphics[width=\textwidth]{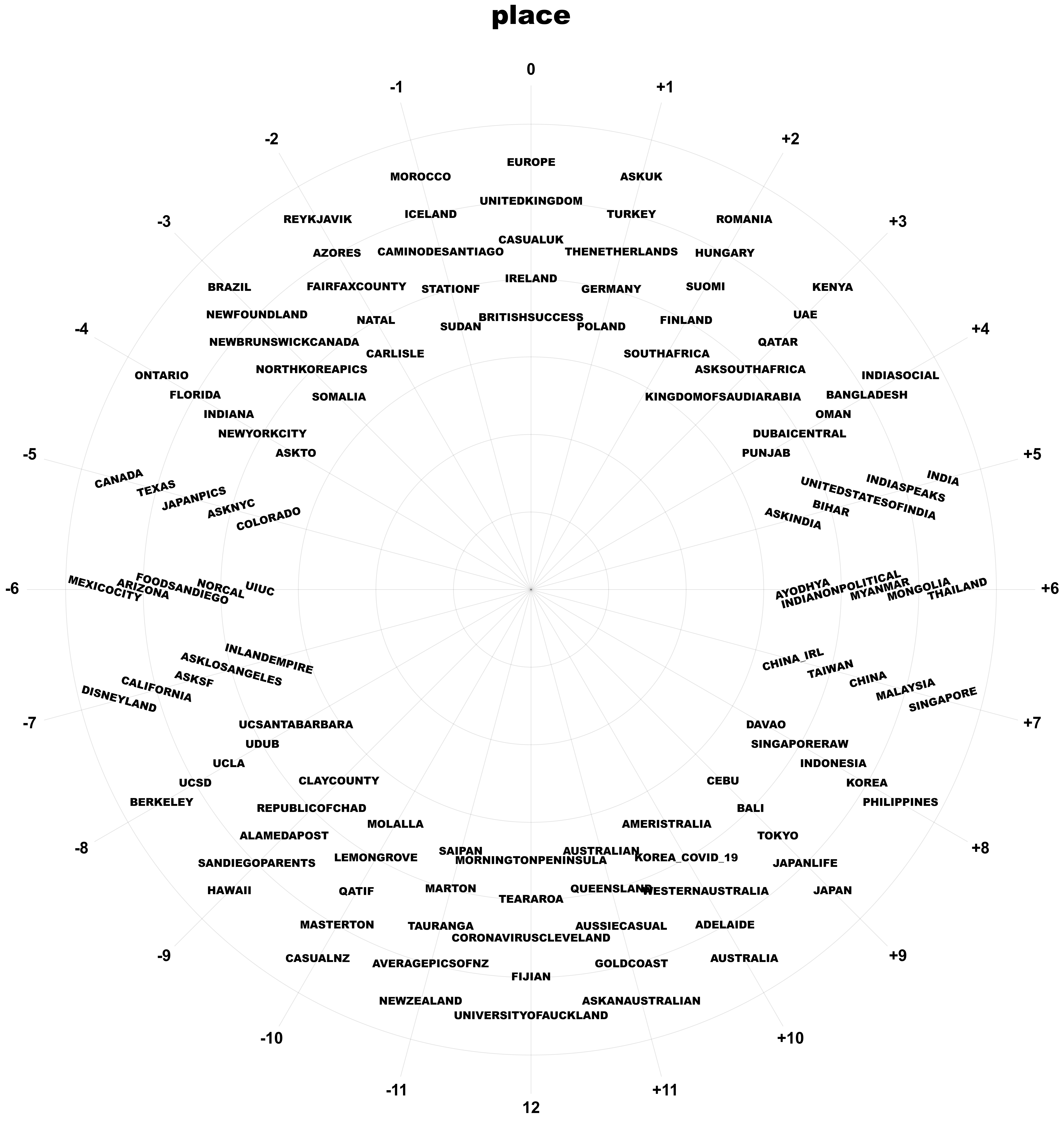}
      \caption{Inferred UTC offsets mapped to semantic categories. The distribution accurately reflects real-world geography, with US cities clustering in negative offsets, European cities near zero, and Asian/Australian locations in positive offsets.}
  \label{fig:semantic_clusters}
\end{figure}  

\begin{figure}[h!]
    \centering
    \includegraphics[width=\textwidth]{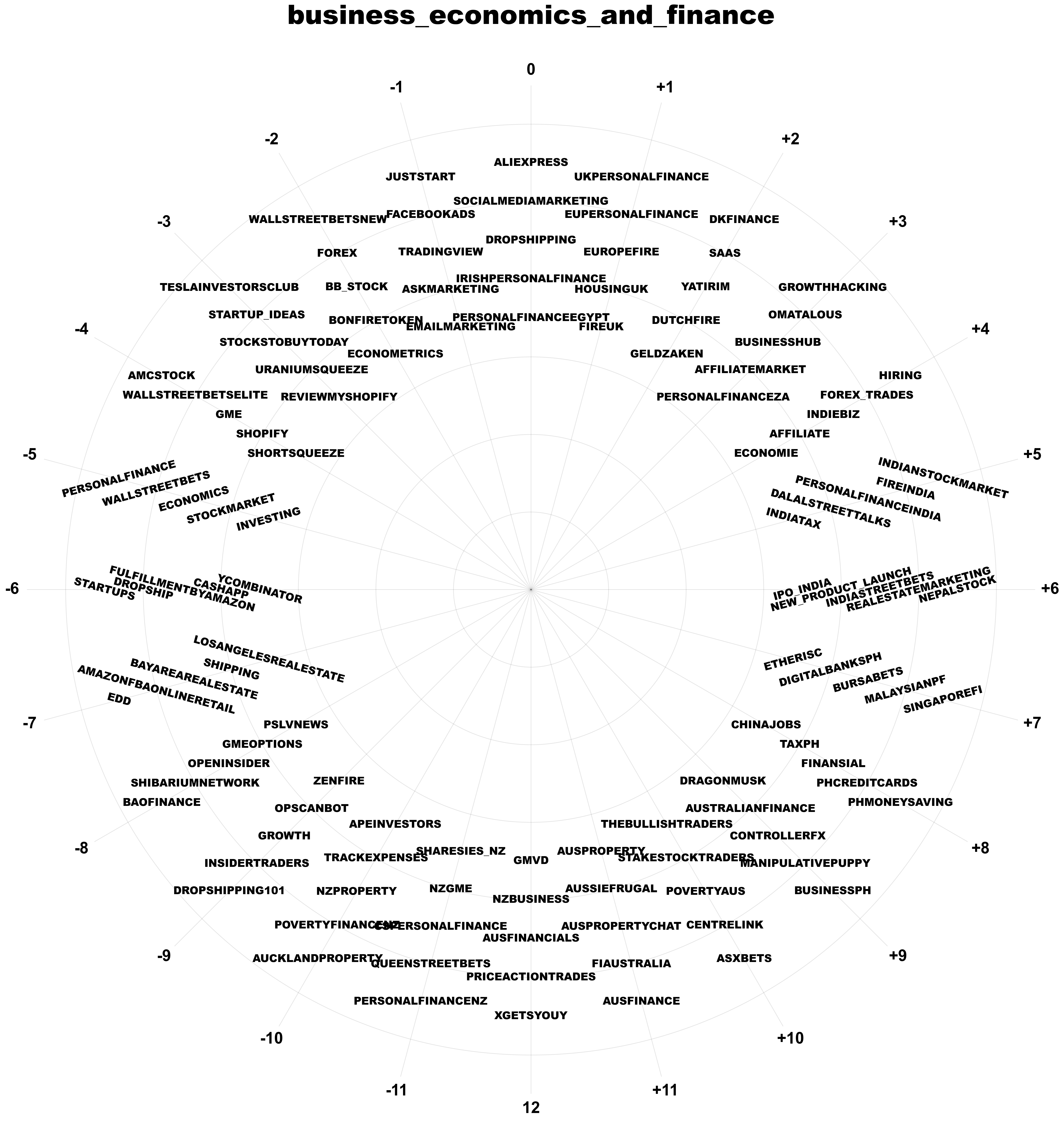}
    
    \caption{Inferred UTC offsets for subreddits in the Business, Economics, and Finance category.}
    \label{fig:subreddit_geo_map_business_economics_and_finance_angled}
  \end{figure}
\begin{figure}[h!]
    \centering
    \includegraphics[width=\textwidth]{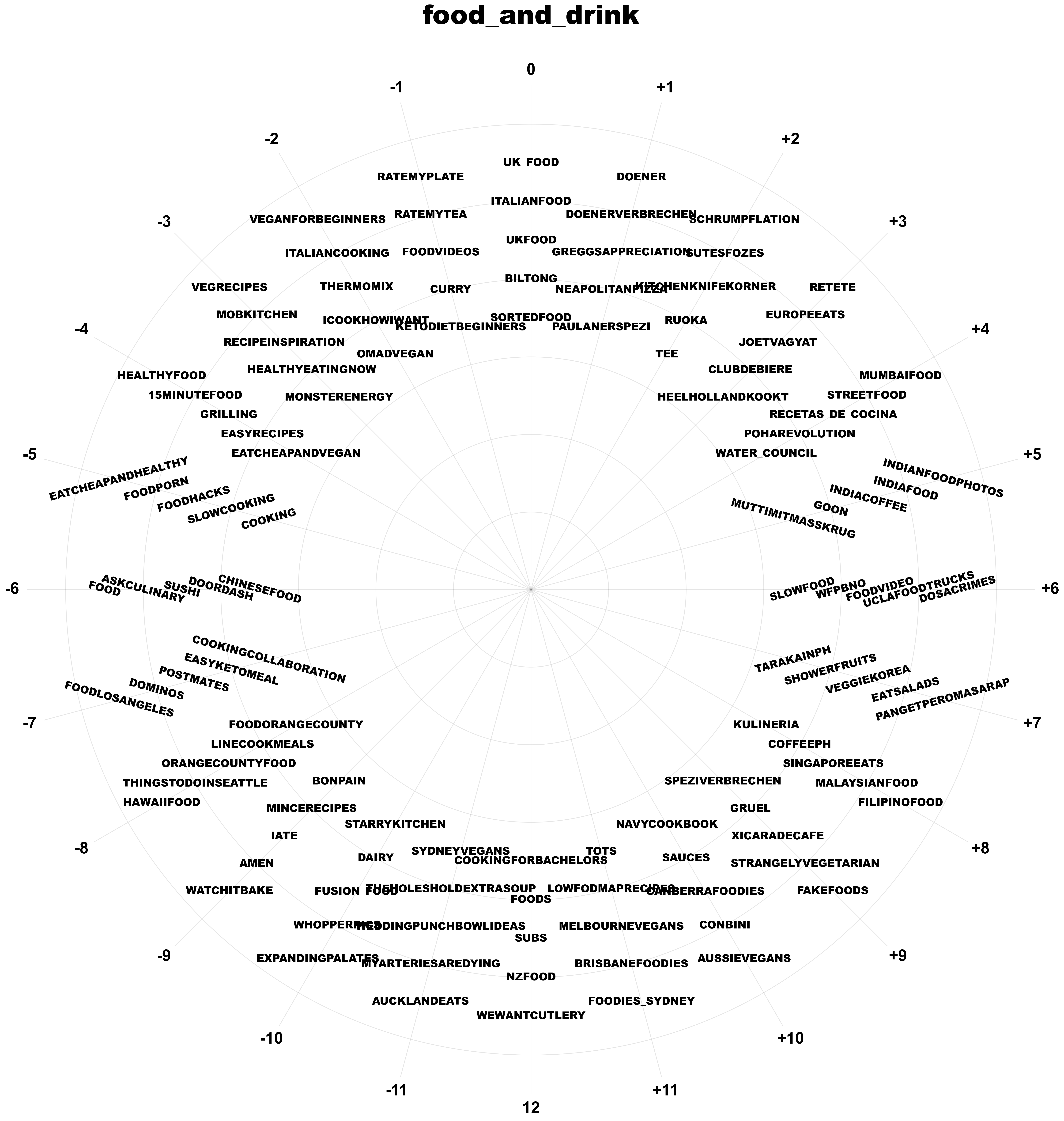}
    \caption{Inferred UTC offsets for subreddits in the Food and Drink category.}
    \label{fig:subreddit_geo_map_food_and_drink_angled}
  \end{figure}

\end{document}